\documentstyle[12pt,epsf]{article}
\newlength{\numlen}

\newlength{\indexlength}

\newcommand{\fr}[2]{{\frac{#1}{#2}}}
\newcommand{\be}{\begin{equation}}
\newcommand{\ee}{\end{equation}}
\newcommand{\ba}{\begin{eqnarray}}
\newcommand{\ea}{\end{eqnarray}}

\newcommand{\rmi}[1]{{\mbox{\scriptsize #1}}}

\newcommand{\nr}[1]{(\ref{#1})}
\newcommand{\msbar}{\overline{\mbox{\rm MS}}}
\newcommand{\Tint}[1]{{\hbox{$\sum$}\!\!\!\!\!\!\int}_{\!\!\!\!#1}}
\newcommand{\lsim}{\raise0.3ex\hbox{$<$\kern-0.75em\raise-1.1ex\hbox{$\sim$}}}
\newcommand{\gsim}{\raise0.3ex\hbox{$>$\kern-0.75em\raise-1.1ex\hbox{$\sim$}}}
\relax
\catcode`\@=11
\def\slash{\@ifnextchar[{\@slash}{\@slash[\z@]}}
\def\@slash[#1]#2{\setbox\z@\hbox{$#2$}\@tempdima\wd\z@\box\z@%
\@tempdimb#1 \advance\@tempdimb-\@tempdima \kern\@tempdimb
\hbox to\@tempdima{\hss\@makeslash\hss}}
\def\@makeslash{$\,/$}
\catcode`\@=12

\begin{document}

\begin{titlepage}
\begin{flushright}
CERN-TH/95-226\\
HU-TFT-95-50\\
IUHET-312\\
August 27, 1995
\end{flushright}
\begin{centering}
\vfill

{\bf GENERIC RULES FOR HIGH TEMPERATURE\\
DIMENSIONAL REDUCTION AND THEIR\\
APPLICATION TO THE STANDARD MODEL}
\vspace{1cm}

K. Kajantie$^{a}$\footnote{kajantie@phcu.helsinki.fi},
M. Laine$^{a}$\footnote{mlaine@phcu.helsinki.fi},
K. Rummukainen$^b$\footnote{kari@trek.physics.indiana.edu} and
M. Shaposhnikov$^c$\footnote{mshaposh@nxth04.cern.ch} \\

\vspace{1cm} {\em $^a$Department of Physics, P.O.Box 9, 00014
University of Helsinki, Finland\\} \vspace{0.3cm} {\em $^b$Indiana
University, Department of Physics, Swain Hall West 117,\\Bloomington
IN 47405 USA\\} \vspace{0.3cm} {\em $^c$Theory Division, CERN,\\
CH-1211 Geneva 23, Switzerland}

\vspace{2cm}

{\bf Abstract}

\vspace{0.5cm}
We formulate the rules for dimensional reduction
of a generic finite temperature gauge theory to a simpler
three-dimensional effective bosonic theory in terms of a matching of
Green's functions in the full and the effective theory,
and present a computation of a generic set
of 1- and 2-loop graphs needed for the application of these rules.
As a concrete application we determine the explicit mapping of the physical
parameters of the standard electroweak theory
to a three-dimensional SU(2)$\times$U(1) gauge-Higgs theory.
We argue that this
three-dimensional theory has a universal character and appears as an
effective theory for many extensions of the Standard Model.
\end{centering}

\vspace{0.3cm}\noindent

\vfill \vfill
\noindent

\end{titlepage}
\section{Introduction}
The properties of matter at high temperature are interesting for a
number of experimental and cosmological applications. QCD at high
temperature and density may be relevant for heavy ion collisions,
while finite temperature phase transitions may play an important role
in the evolution of the universe. In gauge theories, an entirely analytic
perturbative study of the properties of high temperature matter is not
possible due to the so called infrared problem in the thermodynamics of
Yang-Mills fields~\cite{linde}.
A direct way to compute static equilibrium quantities at high
temperature would be to do lattice Monte Carlo
simulations in the 4d high
temperature theory. However, in many interesting cases the use of the
full 4d theory is difficult, if not impossible~\cite{FKRS2}.

These obstacles invoke a demand for a formalism which can solve in
a constructive way the problems mentioned. Since a finite
temperature equilibrium field theory is equivalent to a zero
temperature Euclidean field theory with compact 4th dimension, the
idea of the 4d $\rightarrow$ 3d dimensional reduction
is natural~\cite{G}--\cite{L}.
Dimensional reduction means that some properties of the
{\em equilibrium} high temperature plasma can be derived from a simpler
3d effective theory. The construction of the effective theory
is free of IR problems.
The 3d theory is purely bosonic,
and may then be studied by non-perturbative methods,
such as lattice MC simulations. In fact, the idea of dimensional
reduction has been around for quite
a long time~\cite{G,AP}. However, some concrete analytical results for the
construction of the 3d effective theory have appeared only recently. They
are relevant for the description of the high temperature electroweak phase
transition~\cite{FKRS2},\cite{FKRS1}--\cite{ml2} and high
temperature QCD~\cite{reisz1}--\cite{ay}.

The aim of the present paper is the formulation of the general rules
of dimensional reduction in a constructive way. Namely, we present
a set of 1-loop and 2-loop
Feynman diagrams with the results of their computation which can be
used for dimensional reduction in any gauge field theory.
As an example we construct the 3d effective theory corresponding to
the Standard Model of electroweak interactions.
New elements here in comparison with \cite{FKRS1}
are the inclusion of fermions, and the
direct relation of the parameters of the effective theory to the
physical parameters of the EW theory (the physical Z and
W boson, Higgs particle and top quark masses, the muon lifetime
and the temperature). We also discuss the
strategy for the derivation of the simplest possible effective theory
for typical extensions of the electroweak theory, like the models with
two Higgs doublets, and the Minimal Supersymmetric Standard Model.

The paper is organized as follows. In Section~2 we formulate the
general notion of dimensional reduction and analyse the expansion
parameters involved there. In Section~3 we present the building blocks for
the construction of the effective theory.  Section~4 contains the dimensional
reduction of the Standard Model. In Section~5 we relate the parameters
in the $\msbar$ scheme to the physical parameters, thus completing the
relation of 3d couplings to temperature and observables.
Section~6 is a discussion. We argue there that
the effective theory of most of the extensions of Standard Model is
just the SU(2)$\times$U(1)+Higgs model.

\section{Dimensional reduction}
\label{DR}

The equilibrium properties of matter at high temperatures are related
to Matsubara Green's functions of different field operators.
By the concept of dimensional reduction we mean
that with some accuracy, all the 4d {\em static bosonic}
Green's functions in low energy domain (see below)
can be computed with the help of some effective 3d field theory.
Let us start with useful definitions.

\subsection{Superheavy, heavy and light modes}
\label{accuracy}
In order to define dimensional reduction,
consider a generic renormalizable field theory at high temperature
containing gauge $A_{\mu}$, scalar $\phi$ and fermionic fields
$\psi$,
\be
L={1\over4} F_{\mu\nu}F_{\mu\nu} +
(D_{\mu}\phi)^\dagger(D_{\mu}\phi) + V(\phi) +
g_Y {\bar{\psi}}\phi\psi+\delta L.
\label{lagr}
\ee
Here the group indices are suppressed, $\delta L$
contains the counterterms, and $V(\phi)$ is of the form\footnote{The
analysis of the case when cubic terms are present goes along
the same lines.}
\be
V(\phi)=m_S^2\phi^\dagger\phi+\lambda(\phi^\dagger\phi)^2.
\ee
For power counting
let us assume that $\lambda \sim g^2$ and $g_Y\sim g$, where $g$,
$\lambda$ and $g_Y$ are the gauge, scalar, and Yukawa couplings,
respectively. Write all 4d Matsubara fields in the form
\be
\phi(x,\tau)=\sum_{n=-\infty}^{\infty}\phi_n(x)\exp(i
\omega^b_n \tau),
\ee
\be
\psi(x,\tau)=\sum_{n=-\infty}^{\infty}\psi_n(x)\exp(i
\omega^f_n \tau),
\ee
where $\omega^b_n=2 n \pi T,~ \omega^f_n=(2n+1) \pi T $ are the 3d
tree masses for the bosonic ($\phi_n$) and fermionic ($\psi_n$) 3d
fields. Consider the 1-loop corrections
(for definiteness in the $\msbar$-scheme)
to the masses of the static
modes $\phi_0$ from the modes $\phi_{n\neq 0}$ and $\psi_n$.
In general, they have the form
\be
m_i^2(T)=\gamma_i T^2 + m_i^2,~~ \gamma_i \sim g^2,
\label{mass}
\ee
where $m_i^2$ is the zero-temperature mass of the scalar field
evaluated at some scale $\mu_T^m$ (see below, and~\cite{FKRS1}).
In general, $m^2(T)$ may be matrices, and in the
discussion below we mean the eigenvalues of those.
For the spatial components of the gauge fields $\gamma_i=0,~m_i^2=0$;
for the temporal components of the gauge
fields $\gamma_i\neq 0,~m_i^2=0$;
for the scalar fields $\gamma_i\neq 0,~ m_i^2 \neq 0$.
Now, let us divide the masses into different
categories depending on their magnitude at high temperature.
The 3d masses of all
fermionic modes and all bosonic modes with $n \neq 0$
are proportional to
$\pi T$, and we will call these modes superheavy.
The masses of the temporal components of the gauge fields~$A_0$
are proportional to $g T$, and these modes are called heavy.
The scalar fields can be separated in two different groups.
If $m_i^2$ is different from $-\gamma_i T^2$,
the scalar mass is proportional to $g T$, and the
field corresponding to this mass is ``heavy''.
In the contrary, one may be close to
a tree-level phase transition temperature so that
$\gamma_i T^2$ and $m_i^2$ cancel each other. Then
$m_i^2(T)\sim(g^2T)^2$, and
we call this field ``light''. We denote a generic light
scalar mass by $m_3^2$. All spatial components of
the gauge fields are ``light'' because for
them $\gamma_i=0$.

After these definitions we are ready to explain the
conjecture behind dimensional reduction. Two levels of dimensional
reduction are usually considered. On the first level the
effective theory is constructed for the light and heavy modes
(superheavy modes are ``integrated out''). The second level is the
theory for the light modes only.
In this paper we
require the 3d Lagrangian of the effective theory
to be super-renormalizable, so that
scalar self-interactions are at most quartic.
The super-renormalizable character introduces an
absolute upper bound on the accuracy of the description of
the 4d world by a 3d theory, to be discussed below.

\subsection{Two levels of dimensional reduction}
{\bf The theory for light and heavy modes.}
This theory is valid up
to momenta $k\ll T$, but $k$ may be as large as $g T$.
Consider a (super)renormalizable 3d gauge-Higgs theory with the
Lagrangian
\be
L={1\over4} F_{ij}F_{ij} +
(D_{i}\phi)^\dagger(D_{i}\phi) + V_3(\phi,A_0)+ \fr12 (D_{i}A_0)^2
+\delta L,
\label{3dheavy}
\ee
where $V_3(\phi,A_0)$ is of the form
\be
V_3(\phi,A_0)=m^2\phi^\dagger\phi+\lambda_3(\phi^\dagger\phi)^2+
h_3\phi^\dagger\phi A_0^2+\frac{1}{2}m_D^2A_0^2+
\frac{1}{4}\lambda_A A_0^4.
\ee
The gauge couplings $g_{3}$ have the dimension
GeV$^{1/2}$ and the scalar couplings $\lambda_{3}$
the dimension GeV.
To leading order, the parameter $m_D\sim gT$
is nothing but the Debye mass.
Consider bosonic static
$n$-point one-particle-irreducible Matsubara Green's functions
$G^{(4)}_{n}(\vec{k}_i)$ for the light and heavy fields in the full
4d theory, multiplied by factor $T^{n/2-1}$ to have the
dimension GeV$^{3-n/2}$, and depending on external
3-momenta $\vec{k}_i$. The statement of dimensional reduction is that
there is a mapping of the temperature
and the 4d coupling constants of the underlying theory
to the 3d theory such that the 3d theory gives
the same light and heavy Green's functions as the full 4d theory for
$k\le gT$ up to terms of order $O(g^4)$,
\be
\frac{\Delta G}{G} \sim O(g^4).
\label{accurh}
\ee
Fourth order in $g$ appears from a powercounting estimate
of the contributions of the neglected 6-dimensional operators to
typical Green's functions. For example, the operator $g^6 \phi^2 A_0^4/T^2$
contributes to the 2-point scalar correlator at order $g^6 m_D^2\sim g^8T^2$.
Since the order of magnitude of $m_3^2$ is $g^4 T^2$, the relative
error is $O(g^4)$. The same estimate arises by comparing
the contribution of the neglected operator $g^2 (k^4/T^2)\phi^2$
to the tree-level term $k^2$ at momenta $k\sim gT$.
To reach the accuracy goal~(\ref{accurh}),
the parameters of the 3d theory should be known
with relative uncertainty~$O(g^4)$, which means
1-loop accuracy~[$T(g^2+g^4)$] for the coupling constants,
2-loop accuracy~[$T^2(g^2+g^4)$] for the heavy
masses, and 3-loop accuracy~[$m^2+T^2(g^2+g^4+g^6)$] for the
light scalar masses.

Some comments are now in order.\\
(i) The problem of constructing an
effective 3d theory giving an accuracy better than $O(g^4)$ for {\em all}
Green's functions is far from being trivial (if possible at all).
It is clear, though, that if the theory exists, it must contain 6-dimensional
operators, and the 4d-3d mapping for the light scalar modes must
be done beyond 3-loop level.\\
(ii) Often dimensional reduction is done on the tree-level for
the couplings and 1-loop level
for the masses, i.e., at order $g^2$. This 3d theory
reproduces the 1-loop resummed effective
potential for the Higgs field~\cite{DHLLL,Ca,BFHW}.
However, the relative uncertainty in the mass squared
of the light scalar field is
$\Delta m_3^2/m_3^2\sim O(1)$, since the
tree-level mass term is compensated for by the 1-loop
thermal correction near the phase transition.
Hence ${\Delta G}/{G}\sim O(1)$,
and from the point of view of calculating general
correlators, the theory is useless.
To obtain the minimal useful accuracy $O(g^2)$,
one should go to the 2-loop order $g^4$
in the scalar mass parameter. A more complete $g^4$ calculation,
including 1-loop dimensional reduction [$T(g^2+g^4)$] for
the couplings coupled to the scalar fields, 1-loop
dimensional reduction [$T^2g^2$] for the heavy masses,
and 2-loop dimensional
reduction [$m^2+T^2(g^2+g^4)$] for the scalar mass,
is needed~\cite{FKRS1} to
reproduce the resummed 2-loop effective potential
for the Higgs field~\cite{AE,FH}.
The accuracy $g^4$ corresponds to 1-loop accuracy
in vacuum renormalization, and we will work
with this accuracy throughout this paper. In a weakly
coupled theory, the relative error $O(g^2)\sim g^2/16\pi^2$
of the $g^4$-calculation
is numerically very small. In the Standard Model, the
largest contributions arise from the top quark.\\
(iii) For some quantities, such as the critical temperature and the
observables in  the broken phase, the $g^4$ computation
described in (ii) gives a relative error of order $O(g^4)$.

Consider now the second level of dimensional reduction.

{\bf The theory for light modes only.} This theory is valid up to
momenta $k\ll gT$, but $k$ may be as large as $g^2 T$. The Lagrangian
for this theory is just
\be
L={1\over4} F_{ij}F_{ij} +
(D_{i}\phi)^\dagger(D_{i}\phi) + V_3(\phi),
\label{3dlight}
\ee
where $V_3(\phi)$ is of the form
\be
V_3(\phi)=\bar{m}_3^2\phi^\dagger\phi+\bar{\lambda}_3(\phi^\dagger\phi)^2.
\label{3dlV}
\ee
Only light scalar fields are present.
The effective field theory
can provide the accuracy
\be
\frac{\Delta G}{G} \sim O(g^3).
\label{accurl}
\ee
This estimate arises as follows:
there are neglected 6-dimensional operators
of the form ${g^6 T\phi^6}/{m_{\rm heavy}^3}\sim
{g^3\phi^6}/{T^2}$, contributing to the
two-point scalar correlator at order $g^3m_3^2\sim g^7 T^2$.
This should be compared with $m_3^2\sim g^4T^2$.
Note that in contrast to the integration over the superheavy scale,
odd powers of coupling constants appear, since $m_{\rm heavy} \sim gT$.
To reach the accuracy (\ref{accurl}) one must know
$\bar{\lambda}_3$ in eq.~\nr{3dlV} including corrections
of order $g^4T$ and $\bar{m}_3^2$ including corrections of
order $g^6T^2$.

Comments analogous to the above are applicable to the second level of
dimensional reduction:\\
(ii) To go beyond the accuracy $O(g^3)$, the 6-dimensional operators
must be included in the Lagrangian and light scalar masses must
be computed at least with accuracy $g^7T^2$.\\
(ii) In practice, it is convenient to do the integration over
the heavy scale to the same order in the loop expansion as
the integration over the superheavy scale. This means 1-loop
level [$T(g^2+g^3)$] for the couplings and
2-loop level [$m_3^2+T^2(g^3+g^4)$] for
the scalar mass squared~$\bar{m}_3^2$.
The relative error in the couplings is then $O(g^2)$.
In~$\bar{m}_3^2$, the relative error is $O(g)$, which
is also the relative error in the Green's functions.
Numerically, $O(g)\sim g/4 \pi$ is small in a weakly coupled
theory. Note that since the theory in eq.~\nr{3dheavy} is purely bosonic,
there are no large fermionic corrections.\\
(iii) The procedure described in (ii) provides $O(g^3)$ accuracy in the
critical temperature and the broken phase observables.

Concrete numerical estimates of the accuracy of the effective
field theory depend on the observable and on the details of the
model. Some estimates for the electroweak theory were presented in
\cite{FKRS2,FKRS1,JKP}, and we add some more in Section 5.4.

\subsection{Dimensional reduction by matching}
The definition of dimensional reduction described above provides a
method of mapping the 4d theory on the 3d one. One just writes down
the most general 3d super-renormalizable Lagrangian for the heavy and
light modes, and defines its parameters by matching to a specified
accuracy the 2-, 3-,
and 4-point Green's functions in the 3d effective theory and in the
underlying 4d fundamental theory. The Green's functions to be matched
correspond to those appearing in the 3d Lagrangian. For the
2-point functions one needs the momentum dependent part,
but the 4-point functions may be taken at vanishing external momenta.
Due to gauge invariance, the 3-point
functions are not needed at all.
The scalar Green's functions with vanishing
momenta are most conveniently generated from
an effective potential.

Consider in some more detail the renormalized
2-point function for the light scalar field. In the full 4d
theory, it is of the form
\be
k^2+m_S^2+\Pi(k^2)=k^2+m_S^2+\Pi_3(k^2)+\overline{\Pi}(k^2),\label{4dp}
\ee
and we want to match it to the corresponding function in the 3d theory:
\be
k^2+m_3^2+\Pi_3(k^2). \label{3dres}
\ee
Here $\Pi_3(k^2)$ is the contribution of the light and heavy modes only,
and $\overline{\Pi}(k^2)$ represents all other contributions
(corresponding 2-loop graphs contain at least 2 superheavy
lines)\footnote{To be precise, one must use
resummation to produce the correct
$\Pi_3(k^2)$ in eq.~\nr{4dp};
however, this is not relevant for the present argument.}.
Since there are no IR-problems related to
the integration over the superheavy modes,
$\overline{\Pi}(k^2)$ is analytic in the external momentum~$k^2$,
and can for $k\ll T$ be expanded as
\be
\overline{\Pi}(k^2)=\overline{\Pi}(0)+\overline{\Pi}'(0)k^2+
O(g^2\frac{k^4}{T^2}).\label{pik2}
\ee
Here the $\Pi$'s are of order $g^2$ and if we restrict to $k\le gT$,
the higher-order contributions $O(g^2{k^4}/{T^2})$
are at most of order $g^6T^2$ and can be
neglected.
Assuming that $\overline{\Pi}(0)$ has been calculated
to 2-loop accuracy [$T^2(g^2+g^4)$] and
$\overline{\Pi}'(0)$ to 1-loop accuracy $g^2$, one can
rewrite the right-hand-side of eq.~\nr{4dp} as
\be
[1+\overline{\Pi}'(0)]\Bigl\{
k^2+[m_S^2+\overline{\Pi}(0)][1-\overline{\Pi}'(0)]+\Pi_3(k^2)
\Bigr\},\label{4dres}
\ee
where $\Pi_3(k^2)$ is of order $g^2T m_D\sim g^3T^2$ and
only terms up to order $g^4T^2$ are kept.
The matching of eqs.~\nr{3dres}
and \nr{4dres} can now be carried out by relating the
normalizations of the fields in 3d and 4d through
\be
\phi_\rmi{3d}^2={1\over T}[1+\overline{\Pi}'(0)]\phi_{4d}^2,
\ee
and by relating the masses as
\be
m_3^2=[m_S^2+\overline{\Pi}(0)][1-\overline{\Pi}'(0)],
\ee
which is the order $g^4$ result for $m_3^2$.
The other coupling constants can be fixed
similarly, using the appropriate correlators and
taking always into account the different normalizations
of the fields in 4d and 3d. The 3d theory relevant
for the Standard Model is constructed in this way
in Sec.~\ref{DRinSM}.

The general structure of the
relationships of the 4d and 3d parameters
is determined by
the super-renormalizable character of the 3d theory.
The 4d couplings and masses
are functions of the 4d $\msbar$ parameter~$\mu_4$, but the 3d scalar
and gauge coupling constants are renormalization group (RG) invariant,
since the 3d theory contains only mass divergences. For example, on
the 1-loop level the relationships of the 4d and 3d coupling
constants must have the form
\be
g_3^2 = T[g^2(\mu_4) - \beta_{g^2}\log(\mu_4/c_{g^2} T)],\quad
\lambda_3= T[\lambda(\mu_4) -  \beta_\lambda\log(\mu_4/c_{\lambda}T)],
\ee
where $c_{g^2}$ and $c_{\lambda}$ are
definite fixed functions of physical parameters computable in
perturbation theory (see below),
and the $\beta$'s are the corresponding $\beta$-functions.
The scalar masses in the effective 3d theory, on the other hand,
are not RG-invariant,
but require ultraviolet renormalization on the 2-loop level.
Just dimensionally, the renormalized mass
parameters are of the form
\be
m_3^2(\mu_3)=  {1\over16\pi^2}f_{2m}
\log\frac{\Lambda_m}{\mu_3},
\label{3dmass}
\ee
where $f_{2m}\sim g_3^4$.  For clarity, let us point out
that $\mu_3$ in eq.~\nr{3dmass} is independent of the $\mu_4$
of the 4d theory,
since the bare mass parameter produced by the
dimensional reduction step is RG-invariant.
In Sec.~\ref{blocks} we present a
set of rules, together with a
computation of the necessary Feynman diagrams,
allowing one to define the mapping
of 4d on 3d (at 1-loop level for
coupling constants and 2-loop level for masses)
for an arbitrary gauge theory.

An important comment is now in order. The matching
procedure of dimensional reduction described above
is {\em different} from the initial \cite{G} method
of dimensional reduction which is defined as the sequence of
the following steps: \\
(i) Define a 3d bosonic effective action as
\be
\exp(-S_\rmi{eff})=\int D\psi D\phi_{n \neq 0}\exp(-S),
\label{io}
\ee
where integration over all superheavy modes is performed.\\
(ii)Make a perturbative computation of $S_\rmi{eff}$ and represent it
in
the form
\be
S_\rmi{eff}= c V T^3+\int d^3x L_{\rmi{eff}}(T) + \sum_n
\frac{O_n}{T^n},
\ee
where $L_{\rmi{eff}}(T)$ is a renormalizable 3d effective bosonic
Lagrangian with temperature-dependent constants, $O_n$ are operators
of dimensionality $n$, suppressed by powers of temperature, $c$ is
a number related to the number of degrees of freedom of the theory
and $V$ is the volume of the system.\\
(iii) Drop all the terms $O_n$. The effective action contains then
light and heavy fields. The final step is the integration over the
heavy modes in a way described in (i).
\vspace*{0.5cm}

\begin{figure}[tbh]

\hspace*{0.0cm}
\epsfysize=4cm
\epsffile[100 580 550 700]{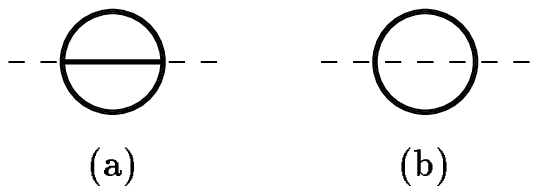}

\caption[a]{\protect
The 2-loop graph (a) relevant for calculating
the dimensionally reduced mass parameter in the
``integration out''-procedure,
and the additional graph (b) needed in the
``matching'' procedure.
The thick lines are superheavy fields,
and the thin dashed lines light fields.}
\label{dr2l}
\end{figure}

The difficulties with the procedure described above,
arising at 2-loop level,  have been pointed
out, e.g.,\ in~\cite{J,Mack}.
The problems are due to steps (ii) and (iii), since
step (i) produces non-local operators which cannot
be expanded in powers of $p^2/T^2$. In terms of
graphs, in the procedure of eq.~\nr{io} the internal
lines of the Feynman diagrams are always superheavy (or heavy). For
example, the only scalar diagram contributing to the scalar mass
renormalization on the 2-loop level is shown in Fig.~\ref{dr2l}.a.
In the Green's function approach the extra graph in
Fig.~\ref{dr2l}.b, containing two superheavy and one light
internal lines appears. As is pointed out in~\cite{FKRS1} this
diagram does not vanish in the high temperature limit,
and therefore, gives a contribution to the 3d mass.
Physically, the reason is that light fields
can have high momenta $p\sim T$ when they interact with
the superheavy fields. The need to include light fields
in the internal lines of many-loop graphs in order to
establish a useful local effective field theory,
is also well known in the context of large-mass expansion in
zero-temperature field theory (see, e.g., \cite{Co}).

When we speak of ``integrating over'' the superheavy or heavy
scale below, we always mean the matching procedure
for the Green's functions described in this Section.

\section{Building blocks for dimensional reduction}
\label{blocks}

In this Section, we give results for the typical diagrams
appearing in the construction of the effective 3d theory.
We account here for the momentum integrations and spin
contractions; the isospin contractions,
combinatorial factors, and coupling constants
relevant for the Standard Model
are added in Sec.~\ref{DRinSM}.
We work in Landau gauge, where the vector propagators
are transversal. The wave function normalization factors
relating the 4d and 3d fields depend
on the gauge condition~\cite{L}, but the final parameters
of 3d theory are gauge-independent at least to the order
in which we are working~\cite{FKRS1,JP}.
Landau-gauge is a convenient choice
since it reduces the number of diagrams considerably:
an external scalar leg with vanishing momentum cannot
directly couple to a vector field, since the vertex is
proportional to the loop momentum,
and hence gives zero when contracted with
the transversal vector propagator.

We will work throughout in Euclidian space.
The conventions for the Euclidian $\gamma$-matrices $\gamma_\mu$
in terms of the Minkowskian matrices $\gamma^\mu$ are
that $\gamma_0=\gamma^0$, $\gamma_i=-i\gamma^i$. The main
properties are $\gamma_\mu^\dagger=\gamma_\mu$,
$\{\gamma_\mu,\gamma_\nu\}=2\delta_{\mu\nu}$,
$\mathop{\rm Tr}\gamma_\mu=0$, $\mathop{\rm Tr} 1=4$.
Due to the relations $t=-i\tau$ and $A_0^M=iA_0^E$
between Minkowskian and Euclidian variables,
the covariant derivative is
$i\gamma^\mu D_\mu^M= -\gamma_\mu D_\mu^E$.
The matrix $\gamma_5$ satisfies
\ba
& & \{\gamma_\mu,\gamma_5\} = 0,\quad\gamma_5^2=1,\quad
\nonumber \\
& & \mathop{\rm Tr}\gamma_5 =
\mathop{\rm Tr}\gamma_5\gamma_\mu\gamma_\nu=0,\quad
\mathop{\rm Tr}
\gamma_5\gamma_\mu\gamma_\nu\gamma_\sigma\gamma_\rho
\propto \epsilon_{\mu\nu\rho\sigma}.
\ea
We define $a_{R,L}=(1\pm\gamma_5)/2$.

The general form of the theory is the following.
There are scalars $\phi$, vector fields~$A^a_\mu$,
ghosts $\eta^a$, and fermions $\psi$. In the symmetric phase, only
the scalar fields have a mass parameter; any mass
parameters are inessential to dimensional reduction, though,
since we assume $m\sim gT$ so that masses contribute
at higher order . The propagators are
\ba
\langle\phi(-p)\phi(p)\rangle & = & \frac{1}{p^2+m_S^2},\quad
\langle A^a_\mu(-p)A^b_\nu(p)\rangle=
\delta^{ab}\frac{\delta_{\mu\nu}-\frac{p_\mu p_\nu}
{p^2}}{p^2}, \nonumber\\
\langle \bar{\eta}^a(p) \eta^b(p)\rangle & = &
-\frac{\delta^{ab}}{p^2},\quad\quad\quad
\langle \bar{\psi}_\alpha(p)\psi_\beta(p)\rangle=
\frac{i\slash{p}_{\beta\alpha}}{p^2}. \label{symprop}
\ea
Defining
\ba
F_{\mu\nu\rho}(p,q,r) & = &
\delta_{\mu\rho}(p_\nu-r_\nu)+
\delta_{\rho\nu}(r_\mu-q_\mu)+
\delta_{\nu\mu}(q_\rho-p_\rho),  \\
G^{abcd}_{\mu\nu\sigma\!\rho} & = &
f^{abe}f^{cde}(\delta_{\mu\sigma}\delta_{\nu\rho}-
\delta_{\mu\rho}\delta_{\nu\sigma})+
(b\leftrightarrow c,\nu\leftrightarrow\sigma)+
(b\leftrightarrow d,\nu\leftrightarrow\rho),\nonumber
\ea
where $f^{abc}$ is antisymmetric,
the theory has the following types of vertices.
The self-interactions of vector fields are due to
vertices of the form
\ba
\mbox{ } & &
igf^{abc}F_{\mu\nu\rho}(p,q,r)A^a_\mu(p)A^b_\nu(q)A^c_\rho(r),\quad
igf^{abc}p_\mu\bar{\eta}^a(p)A^b_\mu(q)\eta^c(r),\nonumber \\
& & g^2G^{abcd}_{\mu\nu\sigma\!\rho}A^a_\mu A^b_\nu A^c_\sigma A^d_\rho.
\label{gaugevs}
\ea
In the actual calculation one only needs the expression
\be
G^{\alpha\alpha cd}_{\mu\nu\sigma\!\rho} =
f^{\alpha ce}f^{\alpha de}(2\delta_{\mu\nu}\delta_{\sigma\!\rho}-
\delta_{\mu\sigma}\delta_{\nu\rho}-
\delta_{\mu\rho}\delta_{\nu\sigma}),
\ee
where $\alpha$ is not summed over,
so that the isospin part separates
for the quartic vertex as it does  for the cubic one.
Fermions interact through vertices of the type
\be
ig\bar{\psi}\gamma_\mu A_\mu a_L\psi,\quad
ig\bar{\psi}\gamma_\mu A_\mu \psi,\quad
g_Y \bar{\psi}\phi\psi,\quad
g_Y \bar{\psi}\gamma_5\phi\psi,\label{fv}
\ee
and the scalar vertices are of the form
\be
\lambda \phi^4,\quad
ig(p_\mu-r_\mu)\phi(p)A_\mu(q)\phi(r),\quad
g^2\phi\phi A_\mu A_\mu.\label{sv}
\ee
In the above formulas, momentum conservation is implied.
The isospin indices are suppressed in eqs.~\nr{fv} and~\nr{sv}.
It turns out that for the calculations in this paper
it is sufficient to treat explicitly only the first and third
vertex in eq.~\nr{fv}, since the other two give
results differing only by trivial numerical coefficients.

In addition to the renormalized vertices,
one needs counterterms. The wave function
counterterms are denoted by
$\delta Z_S=Z_S-1$ (and similarly for the other fields),
where
\be
\phi_B=Z_S^{1/2}\phi,\quad
A_B=Z_V^{1/2}A,\quad
\psi_{L,B}=(Z_F^L)^{1/2}\psi_L,\quad
\psi_{R,B}=(Z_F^R)^{1/2}\psi_R
\ee
and $\psi_{L(R)}=a_{L(R)}\psi$. The only mass counterterm
in the symmetric phase is $\delta m_S^2$.
In the broken phase, the shift in the scalar field
generates mass counterterms for vectors and fermions, as well.
The coupling constant counterterms are denoted by $\delta g^2$,
$\delta \lambda$ and $\delta g_Y$, and are defined by
\ba
& &
g_B^2\phi_B^2 A_B^2=(g^2+\delta g^2)\phi^2A^2,\quad
\lambda_B\phi_B^4=(\lambda+\delta \lambda)\phi^4,\nonumber \\
& & g_{Y,B}\bar{\psi}_B\phi_B\psi_B=(g_Y+\delta g_Y)
\bar{\psi}\phi\psi.
\ea

\subsection{Integration over the superheavy scale}

In this Section
we construct a local 3d effective field
theory which contains the bosonic
$n=0$ Matsubara modes only, and produces
the same static Green's functions as the full 4d theory
with the required accuracy.
As explained in Sec.~\ref{DR},
the recipe is to first identify the
general structure of the effective theory, and then to
compare static correlators calculated from the 3d
and 4d theories. The structure of the effective theory
differs from the tree-level action for $n=0$ modes
in the 4d theory in that the absence of Lorentz symmetry
allows the temporal components of the gauge fields to
develop mass terms and quartic self-interactions.
At 1-loop level, the construction of the 3d theory
proceeds simply by calculating the effect of
fermions and $n\neq 0$ bosons to two-, \mbox{three-,} and
four-point correlators of the static modes.
At 2-loop level, there can be $n=0$ modes in the
loops, as well, and hence one must carefully compare the
correlators in the two theories.
In Sec.~\ref{3dwf}
we calculate how the 3d fields are related to the 4d fields,
in Sec.~\ref{3dgs} we compute the effective couplings of the gauge sector,
in Sec.~\ref{fundam} we address the fundamental scalar sector, and in
Sec.~\ref{adjoint} we study the adjoint scalar sector, which is
composed of the temporal
components of the gauge fields.

\subsubsection{Notation and basic integrals}

To give results for the diagrams appearing in the integration
over the superheavy fields, we use the following notation:
\ba
\Tint{p} & = & T\sum_n\int\frac{d^dp}{(2\pi)^d},\quad
\Tint{p}' =  T\sum_{n\neq 0}\int\frac{d^dp}{(2\pi)^d},\quad
d=3-2\epsilon,\nonumber \\
p_b & = & (\omega_n^b,\vec{p})
,\quad p_f=(\omega_n^f,\vec{p}),\quad
\omega^b_n=2 n \pi T,\quad \omega^f_n=(2n+1) \pi T,\quad
k \equiv (0,\vec{k}), \nonumber \\
\imath_\epsilon & = &\ln\frac{\mu^2}{T^2}+2 \gamma_E-2\ln 2-
2\frac{\zeta'(2)}{\zeta(2)},\quad
c=\frac{1}{2}\biggl[\ln \frac{8\pi}{9}+\frac{\zeta'(2)}{\zeta(2)}-
2\gamma_E]\approx -0.348725, \nonumber \\
{c_B} & = & \ln (4\pi)-\gamma_E \approx
1.953808,\quad c_F=c_B-2\ln 2\approx
0.567514, \nonumber \\
L_b & = & \ln\frac{\mu^2}{T^2}-2 c_B,\quad
L_f=\ln\frac{\mu^2}{T^2}-2 c_F,\quad
\frac{1}{\epsilon_b}=\frac{1}{\epsilon}+L_b,\quad
\frac{1}{\epsilon_f}=\frac{1}{\epsilon}+L_f.\label{nabi}
\ea
The theory is regularized in the $\overline{\rm MS}$-scheme,
$\mu$ is the corresponding scale parameter.

The basic integrals appearing in 1-loop
integration over the superheavy modes are the following.
The fermionic and bosonic tadpole integrals, to the
accuracy they are needed, are~\cite{AE}
\ba
I_b'(m) & = & \Tint{p_b}' \frac{1}{p^2+m^2}=\mu^{-2\epsilon}
\biggl[\frac{T^2}{12}(1+\epsilon\imath_\epsilon)-
\frac{m^2}{16\pi^2}\biggl(\frac{1}{\epsilon}+L_b\biggr)
\biggr], \label{Ib} \\
I_f(m) & = & \Tint{p_f} \frac{1}{p^2+m^2}=\mu^{-2\epsilon}
\biggl[-\frac{T^2}{24}\Bigl[1+\epsilon
(\imath_\epsilon-2\ln 2)\Bigr]-
\frac{m^2}{16\pi^2}\biggl(\frac{1}{\epsilon}+L_f\biggr) \label{If}
\biggr].
\ea
Taking derivatives with respect to mass squared and temperature
in eqs.~\nr{Ib}, \nr{If}, one can derive other integrals.
In the end one can put the masses in the propagators
to zero, since the integrals
over superheavy modes are analytic in the mass parameters, and
hence the effect of higher orders is suppressed by $m^2/T^2$.
The dependence on external momenta is likewise
analytic, and can be expanded in $k^2/T^2$.
Since all the parameters of the
effective theory are at most of order $gT$,
higher order contributions in $k^2/T^2$
can only produce contributions suppressed
by coupling constants.
The masses will play a role only in Sec.~\ref{fundam},
where we calculate integrals over superheavy modes
not directly, but by using the effective potential;
the needed integrals are given there.
The required massless integrals are
\ba
B_b'& \equiv & \Tint{p_b}'\frac{1}{(p^2)^2}=
\frac{1}{16\pi^2}\frac{1}{\epsilon_b}, \nonumber \\
B_b'(k) & = & \Tint{p_b}'\frac{1}{p^2(p+k)^2} =
\frac{1}{16\pi^2}\frac{1}{\epsilon_b}
\biggl[1+{\cal O}\Bigl(\frac{k^2}{T^2}\Bigr)\biggr], \nonumber \\
J^b_{\alpha\beta}
& \equiv &
\Tint{p_b}'\frac{p_\alpha p_\beta}{p^2(p+k)^2}
-\Tint{p_b}'\frac{p_\alpha p_\beta}{(p^2)^2}, \nonumber \\
J^b_{00} & = & - \frac{k^2}{16\pi^2}
\biggl(\frac{1}{12\epsilon_b}+\frac{1}{6}\biggr), \nonumber \\
J^b_{ij} & = & - \biggl(\delta_{ij}-\frac{k_ik_j}{k^2}\biggr)
\frac{k^2}{16\pi^2}
\biggl(\frac{1}{12\epsilon_b}\biggr)+
\frac{k_ik_j}{16\pi^2}
\biggl(\frac{1}{4\epsilon_b}\biggr),
\label{integrals} \\
K^b_{\alpha\beta}
& \equiv &
\Tint{p_b}'\frac{p_\alpha p_\beta}{(p^2)^2(p+k)^2}, \nonumber \\
K^b_{00} & = & \frac{1}{16\pi^2}
\biggl(\frac{1}{4\epsilon_b}+\frac{1}{2}\biggr), \nonumber \\
K^b_{ij} & = &
\frac{\delta_{ij}}{16\pi^2}
\biggl(\frac{1}{4\epsilon_b}\biggr), \nonumber \\
L^b_0 & \equiv & \Tint{p_b}'\frac{p_0^4}{(p^2)^4}=
\frac{1}{16\pi^2}
\biggl(\frac{1}{8\epsilon_b}+\frac{1}{3}\biggr).
\nonumber
\ea
Here we did not write
$\mu^{-2\epsilon}$
explicitly and neglected the higher-order
contributions in~$k^2/T^2$.
For the fermionic case one simply replaces $1/\epsilon_b$
by $1/\epsilon_f$ everywhere in eq.~\nr{integrals}.

\subsubsection{Wave function normalization}
\label{3dwf}

Let us calculate how the 3d fields are
related to the 4d fields. This is to be done on 1-loop level.
In practice, one has to calculate the contribution of the superheavy modes
to the momentum-dependent part of the two-point correlator
of the light and heavy modes. Indeed, the contribution of
the light and heavy modes
is the same in the full theory and the effective theory, whereas
the contribution of the superheavy modes
can be produced in the effective theory
only by a different normalization of the fields.

\begin{figure}[tbh]

\hspace*{0.0cm}
\epsfysize=7cm
\epsffile[100 500 500 700]{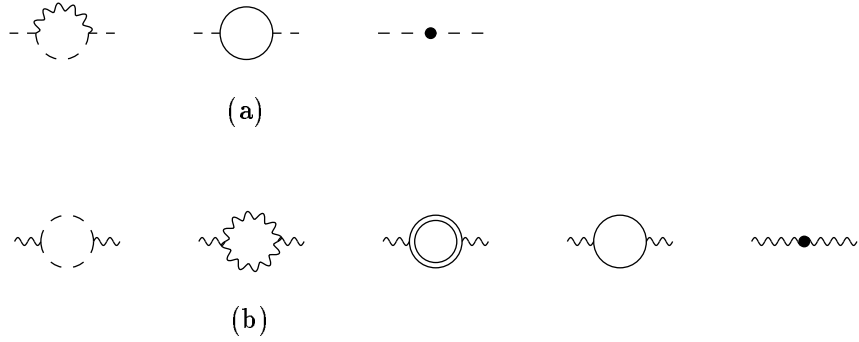}

\caption[a]{\protect
The diagrams needed for the dimensional reduction of
(a) the wave function~$\phi$, and
(b) the wave functions $A_i$ and $A_0$.
Dashed line is a scalar propagator,
wiggly line a vector propagator,
double line a ghost propagator, and
solid line a fermion propagator.
The bare blob indicates the
wave function counterterm.}
\label{drpi}
\end{figure}

The generic diagrams needed for the scalar correlator,
and for the temporal and spatial components of the
vector correlator, are shown in Figs.~\ref{drpi}.a and~\ref{drpi}.b.
To determine the wave function normalization factor,
one needs only the parts proportional to $k^2$ from these diagrams.
We identify the diagram by the types of propagators that appear
in it, S, V, F, and $\eta$ denoting the scalar, vector, fermion
and ghost propagators. Counterterm contributions are denoted by CT.
After some simple algebra one gets
for the diagrams of Fig.~\ref{drpi}.a the results
\ba
{\cal Z}^{\phi}_{\rm CT} & = &  k^2 \delta Z_S,\label{zfct}\\
{\cal Z}^{\phi}_{\rm SV} & = &
\Tint{p_b}' \frac{(2 k_\mu+p_\mu)(2k_\nu+p_\nu)
\Bigl(\delta_{\mu\nu}-\frac{p_\mu p_\nu}{p^2}\Bigr)}
{p^2[(p+k)^2+m_S^2]} \nonumber \\
& \Rightarrow & 4k^2B_b'-4k_ik_j K^b_{ij}
=\frac{k^2}{16\pi^2}\frac{3}{\epsilon_b}, \label{zfsv}\\
{\cal Z}^{\phi}_{\rm FF} & = &
\Tint{p_f} \frac{\mathop{\rm Tr}(i\slash{p})[i\slash[-0.4mm]{(p+k)}]}
{p^2(p+k)^2} \nonumber \\
& \Rightarrow & 2k^2B_f'=\frac{k^2}{16\pi^2}\frac{2}{\epsilon_f}.
\label{zfff}
\ea
When the correct coefficients are taken into account,
the counterterm contribution
${\cal Z}^{\phi}_{\rm CT}$ cancels the $1/\epsilon$-parts from
the two other contributions,
since there is no wave-function renormalization
in the 3d theory.
The remaining $L_b$- and $L_f$-terms determine the relation
of the 3d fields to the 4d fields.
Explicit expressions for the EW theory are given in Sec.~\ref{DRinSM}.

For the vector correlator, the spatial and temporal components
have to be calculated separately.
For the spatial components, one only needs to calculate the transversal
part, and hence the longitudinal part is not displayed below.
The symbols $J^{b(T)}_{ij}$, $K^{b(T)}_{ij}$ mean
the transversal parts of $J^{b}_{ij}$, $K^{b}_{ij}$ in eq.~\nr{integrals}.
The diagrams in Fig.~\ref{drpi}.b give
\ba
{\cal Z}^{A_0}_{\rm CT} & = & k^2\delta Z_V,\label{za0ct}\\
& & \nonumber \\
{\cal Z}^{A_i}_{\rm CT} & = & k^2
\biggl(\delta_{ij}-\frac{k_ik_j}{k^2}\biggr)\delta Z_V,\\
& & \nonumber \\
{\cal Z}^{A_0}_{\rm SS} & = &
\Tint{p_b}'\frac{(2p_0)(2p_0)}{[p^2+m_S^2]
[(p+k)^2+m_S^2]} \nonumber \\
& \Rightarrow & 4J^b_{00}=\frac{k^2}{16\pi^2}
\biggl(-\frac{1}{3\epsilon_b}-\frac{2}{3}\biggr),\label{za0ss}\\
& & \nonumber \\
{\cal Z}^{A_i}_{\rm SS} & = &
\Tint{p_b}'\frac{(2p_i+k_i)(2p_j+k_j)}{[p^2+m_S^2]
[(p+k)^2+m_S^2]} \nonumber \\
& \Rightarrow & 4J^{b(T)}_{ij}=
\biggl(\delta_{ij}-\frac{k_ik_j}{k^2}\biggr)
\frac{k^2}{16\pi^2}
\biggl(-\frac{1}{3\epsilon_b}\biggr),\label{zaiss}\\
& & \nonumber \\
{\cal Z}^{A_0}_{\rm \eta\eta} & = &
\Tint{p_b}'\frac{p_0^2}{p^2
(p+k)^2} \nonumber \\
& \Rightarrow & J^{b}_{00}=
\frac{k^2}{16\pi^2}
\biggl(-\frac{1}{12\epsilon_b}-\frac{1}{6}\biggr),\label{za0ee}\\
& & \nonumber \\
{\cal Z}^{A_i}_{\rm \eta\eta} & = &
\Tint{p_b}'\frac{p_i(p_j+k_j)}{p^2
(p+k)^2} \nonumber \\
& \Rightarrow & J^{b(T)}_{ij}=
\biggl(\delta_{ij}-\frac{k_ik_j}{k^2}\biggr)
\frac{k^2}{16\pi^2}
\biggl(-\frac{1}{12\epsilon_b}\biggr), \\
& & \nonumber \\
{\cal Z}^{A_0}_{\rm FF} & = &
\Tint{p_f}\frac{\mathop{\rm Tr}
[i\slash{p}\gamma_0 a_L][i\slash[-0.4mm]{(p+k)}\gamma_0 a_L]}
{p^2(p+k)^2} \nonumber \\
& = & -2\; \Tint{p_f}\frac{
2p_0^2-\delta_{00}(p^2+p\cdot k)}
{p^2(p+k)^2} \nonumber \\
& \Rightarrow & -4J^f_{00}-k^2B_f'=\frac{k^2}{16\pi^2}
\biggl(-\frac{2}{3\epsilon_f}+\frac{2}{3}\biggr),
\label{za0ff}\\
& & \nonumber \\
{\cal Z}^{A_i}_{\rm FF} & = &
\Tint{p_f}\frac{\mathop{\rm Tr}
[i\slash{p}\gamma_i a_L][i\slash[-0.4mm]{(p+k)}\gamma_j a_L]}
{p^2(p+k)^2} \nonumber \\
& = & -2\; \Tint{p_f}\frac{
2p_ip_j+p_ik_j+p_jk_i-\delta_{ij}(p^2+p\cdot k)}
{p^2(p+k)^2} \nonumber \\
& \Rightarrow & -4J^{f(T)}_{ij}-
\biggl(\delta_{ij}-\frac{k_ik_j}{k^2}\biggr)
k^2B_f' =
\biggl(\delta_{ij}-\frac{k_ik_j}{k^2}\biggr)
\frac{k^2}{16\pi^2}
\biggl(-\frac{2}{3\epsilon_f}\biggr). \label{zaiee}
\ea
To give the two remaining
contributions ${\cal Z}^{A_0}_{\rm VV}$ and ${\cal Z}^{A_i}_{\rm VV}$,
we note that
\be
{\cal Z}^{A_0}_{\rm VV} = \Tint{p_b}'\frac{M_{00}}{p^2(p+k)^2},\quad
{\cal Z}^{A_i}_{\rm VV} = \Tint{p_b}'\frac{M_{ij}}{p^2(p+k)^2},
\ee
where, apart from terms proportional to $k_\mu$,~\cite{PT}
\ba
M_{\mu\nu}
& = &
\biggl(\delta_{\alpha\beta}-\frac{p_\alpha p_\beta}{p^2}\biggr)
\biggl(\delta_{\sigma\rho}-\frac{(p+k)_\sigma (p+k)_\rho}{(p+k)^2}\biggr)
F_{\mu\alpha\sigma}(k,p,-p-k)F_{\nu\beta\rho}(k,p,-p-k) \nonumber \\
& \Rightarrow & \Bigl[p^2+(p+k)^2+4k^2\Bigr]\delta_{\mu\nu}+
(10-8\epsilon)p_\mu p_\nu \nonumber \\
& & -\frac{2}{p^2}
\Bigl[(p^2+2p\cdot k)^2\delta_{\mu\nu}-
(p^2+2p\cdot k-k^2)p_\mu p_\nu \Bigr] +
\frac{1}{p^2(p+k)^2}\Bigl[k^4p_\mu p_\nu\Bigr].\label{Muv}
\ea
Here we utilized the symmetry of the integrand in the change $p\to -p-k$.
The results for the $k^2$-terms can then be seen to be
\ba
{\cal Z}^{A_0}_{\rm VV}
& \Rightarrow &
4k^2 B_b'+(10-8\epsilon)J^b_{00}-
2\Bigl[-k^2B_b'+2k^2K^b_{00}\Bigr]=
\frac{k^2}{16\pi^2}
\biggl(\frac{25}{6}\frac{1}{\epsilon_b}-3\biggr),\label{za0vv}\\
& & \nonumber \\
{\cal Z}^{A_i}_{\rm VV}
& \Rightarrow &
4k^2 B_b'\biggl(\delta_{ij}-\frac{k_ik_j}{k^2}\biggr)
+(10-8\epsilon)J^{b(T)}_{ij}-
2\Bigl[-k^2B_b'\biggl(\delta_{ij}-\frac{k_ik_j}{k^2}\biggr)\nonumber \\
& & +2k^2K^{b(T)}_{ij}\Bigr] =
\biggl(\delta_{ij}-\frac{k_ik_j}{k^2}\biggr)
\frac{k^2}{16\pi^2}
\biggl(\frac{25}{6}\frac{1}{\epsilon_b}+\frac{2}{3}\biggr).\label{zaivv}
\ea
The constant $2/3$ in eq.~\nr{zaivv}
comes from $-8\epsilon J^{b(T)}_{ij}$.
When all the contributions are summed together
with the correct coefficients,
the counterterm contributions ${\cal Z}^{A}_{\rm CT}$
again cancel the $1/\epsilon$-parts.

\subsubsection{The couplings of the gauge sector}
\label{3dgs}

\begin{figure}[tbh]

\hspace*{2.0cm}
\epsfysize=8.5cm
\epsffile[100 440 500 700]{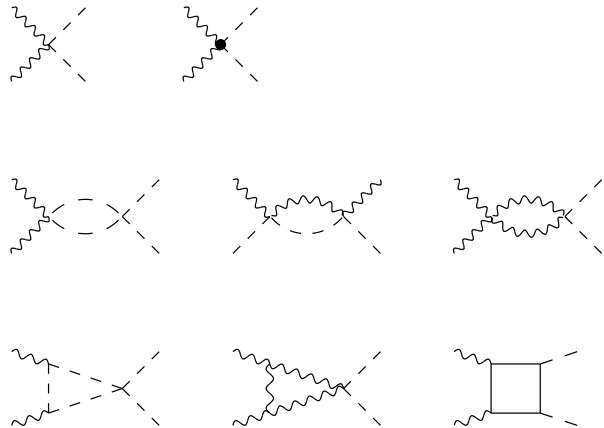}

\caption[a]{\protect
The diagrams needed for calculating
the 3d coupling constants $g_3^2$ and $h_3$.}
\label{drg3}
\end{figure}

To calculate the couplings of the gauge sector,
one has to study some vertex to which
the gauge fields couple. The spatial gauge fields feel only
one coupling constant~$g_3^2$ due to gauge invariance. The interaction
of the temporal components of the gauge fields with the other scalar
fields is not protected by gauge invariance, and hence the
corresponding couplings may differ from $g_3^2$. We calculate
the couplings related to the gauge fields from
a four-point correlator, since
the external momenta may then be assumed to be zero.
In practice, it is most convenient to choose
the $(\phi\phi A A)$-correlator,
since then one gets the two couplings related to
the $(\phi\phi A_iA_j)$- and $(\phi\phi A_0A_0)$-vertices from
almost the same calculations.
The diagrams needed are shown in Fig.~\ref{drg3}.
The results are (${\cal G}_{0}$ is the tree-level contribution)
\ba
{\cal G}^{A_0}_{0} & = &
{\cal G}^{A_i}_{0} = g^2, \label{ga00}
\\ \nonumber \\
{\cal G}^{A_0}_{\rm CT} & = &
{\cal G}^{A_i}_{\rm CT} = \delta g^2,
\\ \nonumber \\
{\cal G}^{A_0}_{\rm SS} & = &
\Tint{p_b}'\frac{1}{(p^2)^2}=B_b'=\frac{1}{16\pi^2}\frac{1}{\epsilon_b},
\\ \nonumber \\
{\cal G}^{A_i}_{\rm SS} & = &
\Tint{p_b}'\frac{\delta_{ij}}{(p^2)^2}=\delta_{ij}B_b'=
\frac{\delta_{ij}}{16\pi^2}\frac{1}{\epsilon_b}, \label{gaiss}
\\ \nonumber \\
{\cal G}^{A_0}_{\rm SV} & = &
\Tint{p_b}'\frac{
\Bigl(\delta_{00}-\frac{p_0^2}{p^2}\Bigr)}{(p^2)^2}
\nonumber \\
& & = B_b'- K^b_{00}
=\frac{1}{16\pi^2}
\biggl(\frac{3}{4}\frac{1}{\epsilon_b}-\frac{1}{2}\biggr),
\\ \nonumber \\
{\cal G}^{A_i}_{\rm SV} & = &
\Tint{p_b}'\frac{
\Bigl(\delta_{ij}-\frac{p_ip_j}{p^2}\Bigr)}{(p^2)^2}
\nonumber \\
& & = \delta_{ij}B_b'- K^b_{ij}
=\frac{\delta_{ij}}{16\pi^2}
\biggl(\frac{3}{4}\frac{1}{\epsilon_b}\biggr),
\\ \nonumber \\
{\cal G}^{A_0}_{\rm VV} & = &
\Tint{p_b}'\frac{
\Bigl(\delta_{\alpha\mu}-\frac{p_\alpha p_\mu}{p^2}\Bigr)
\Bigl(\delta_{\alpha\nu}-\frac{p_\alpha p_\nu}{p^2}\Bigr)}{(p^2)^2}
\Bigl(2\delta_{\mu\nu}\delta_{00}-2\delta_{\mu 0}\delta_{\nu 0}\Bigr)
\nonumber \\
& & = 4(1-\epsilon) B_b'+2 K^b_{00}
=\frac{1}{16\pi^2}
\biggl(\frac{9}{2}\frac{1}{\epsilon_b}-3\biggr),
\\ \nonumber \\
{\cal G}^{A_i}_{\rm VV} & = &
\Tint{p_b}'\frac{
\Bigl(\delta_{\alpha\mu}-\frac{p_\alpha p_\mu}{p^2}\Bigr)
\Bigl(\delta_{\alpha\nu}-\frac{p_\alpha p_\nu}{p^2}\Bigr)}{(p^2)^2}
\Bigl(2\delta_{\mu\nu}\delta_{ij}-\delta_{\mu i}\delta_{\nu j}
-\delta_{\mu j}\delta_{\nu i}\Bigr)
\nonumber \\
& & = 4(1-\epsilon) B_b'\delta_{ij}+2 K^b_{ij}
=\frac{\delta_{ij}}{16\pi^2}
\biggl(\frac{9}{2}\frac{1}{\epsilon_b}-4\biggr),
\\ \nonumber \\
{\cal G}^{A_0}_{\rm SSS} & = &
\Tint{p_b}'\frac{(2p_0)^2}{(p^2)^3} =
4 K^b_{00}=\frac{1}{16\pi^2}
\biggl(\frac{1}{\epsilon_b}+2\biggr),
\\ \nonumber \\
{\cal G}^{A_i}_{\rm SSS} & = &
\Tint{p_b}'\frac{(2p_i)(2p_j)}{(p^2)^3} =
4 K^b_{ij}=\frac{\delta_{ij}}{16\pi^2}
\biggl(\frac{1}{\epsilon_b}\biggr), \label{gaisss}
\\ \nonumber \\
{\cal G}^{A_0}_{\rm VVV} & = &
\Tint{p_b}'
\frac{
\Bigl(\delta_{\alpha\mu}-\frac{p_\alpha p_\mu}{p^2}\Bigr)
\Bigl(\delta_{\alpha\nu}-\frac{p_\alpha p_\nu}{p^2}\Bigr)
\Bigl(\delta_{\sigma\rho}-\frac{p_\sigma p_\rho}{p^2}\Bigr)}
{(p^2)^3}
F_{0\mu\sigma}(0,p,-p)F_{0\nu\rho}(0,p,-p) \nonumber \\
& = &
4 (3-2\epsilon)\Tint{p_b}'
\frac{p_0^2}{(p^2)^3}=
4 (3-2\epsilon) K^b_{00}=
\frac{1}{16\pi^2}
\biggl(\frac{3}{\epsilon_b}+4\biggr),
\\ \nonumber \\
{\cal G}^{A_i}_{\rm VVV} & = &
\Tint{p_b}'
\frac{
\Bigl(\delta_{\alpha\mu}-\frac{p_\alpha p_\mu}{p^2}\Bigr)
\Bigl(\delta_{\alpha\nu}-\frac{p_\alpha p_\nu}{p^2}\Bigr)
\Bigl(\delta_{\sigma\rho}-\frac{p_\sigma p_\rho}{p^2}\Bigr)}
{(p^2)^3}
F_{i\mu\sigma}(0,p,-p)F_{j\nu\rho}(0,p,-p) \nonumber \\
& = &
4 (3-2\epsilon)\Tint{p_b}'
\frac{p_ip_j}{(p^2)^3}=
4 (3-2\epsilon) K^b_{ij}=
\frac{\delta_{ij}}{16\pi^2}
\biggl(\frac{3}{\epsilon_b}-2\biggr),
\\ \nonumber \\
{\cal G}^{A_0}_{\rm FFFF} & = &
\Tint{p_f}\frac{1}{(p^2)^4}
\mathop{\rm Tr} [(i\slash{p})(i\slash{p})(i\slash{p}\gamma_0 a_L)
(i\slash{p} \gamma_0 a_L)] \nonumber \\
& = &
2\;\Tint{p_f}\frac{2p_0^2-p^2\delta_{00}}{(p^2)^3}
=4K^f_{00}-2B_f'
=\frac{1}{16\pi^2}
\biggl(-\frac{1}{\epsilon_f}+2\biggr),
\\ \nonumber \\
{\cal G}^{A_i}_{\rm FFFF} & = &
\Tint{p_f}\frac{1}{(p^2)^4}
\mathop{\rm Tr} [(i\slash{p})(i\slash{p})(i\slash{p}\gamma_i a_L)
(i\slash{p} \gamma_j a_L)] \nonumber \\
& = &
2\;\Tint{p_f}\frac{2p_ip_j-p^2\delta_{ij}}{(p^2)^3}
=4K^f_{ij}-2B_f'\delta_{ij}
=\frac{\delta_{ij}}{16\pi^2}
\biggl(-\frac{1}{\epsilon_f}\biggr). \label{gaiffff}
\ea
The counterterm contributions
${\cal G}^{A_0}_{\rm CT}$, ${\cal G}^{A_i}_{\rm CT}$
cancel the $1/\epsilon$-parts from the other
contributions, since there is no
coupling constant renormalization in the 3d theory.
The final result for the 3d couplings
then consists of the tree-level result
corrected by logarithmic terms and constants.

\subsubsection{The couplings of the fundamental scalar sector}
\label{fundam}

To derive the 3d mass and self-coupling of the $\phi$-field,
one has to calculate the effect of the superheavy modes on
the two- and four-point scalar correlators
with vanishing external momenta.
These contributions can most easily be derived by calculating the
effective potential $V(\varphi)$ for the scalar field,
and extracting from it the part coming from the superheavy modes.
The effective potential contains the one-particle-irreducible
Green's functions $G_n$ at vanishing external momenta
through $V(\varphi)=\sum_n(1/n!)G_n\varphi^n$,
so that the terms
quadratic and quartic in $\varphi$ give the two- and four-point correlators.
The usefulness of $V(\varphi)$ lies in the fact
that the combinatorial factors associated with it
are simpler than those associated
with a direct evaluation of Feynman diagrams.

Since
the superheavy modes do not suffer from IR-problems,
their contribution to the effective potential
is analytic in the mass parameters appearing in the
propagators. In contrast to the direct evaluation of superheavy
contributions in the previous sections, the masses cannot here
be neglected, though, but are quite essential:
the mass parameters depend quadratically
on the shifted field $\varphi$, so that terms of the
form $T^2m^2$, $m^4$ determine the
two- and four-point scalar correlators. To get
the quartic coupling, it is enough to extract the $m^4$-term from
the 1-loop effective potential. For the mass
parameter $m_3^2$, however, one needs a 2-loop calculation.

Let us note that to get the correct result to
order $g^4$ for $V(\varphi)$
actually requires resummation~\cite{AE}. This can be done by
adding and subtracting from the Lagrangian the 1-loop
thermal mass terms
\be
\overline{\Pi}_{\phi}(0)\phi(0,\vec{k})\phi(0,-\vec{k}),\quad
\frac{1}{2}\overline{\Pi}_{A_0}(0)A_0(0,\vec{k})A_0(0,-\vec{k}),
\label{TCT}
\ee
where the bar indicates that only contributions from
the superheavy modes are included.
The terms added to $\cal L$ with plus-signs are treated as tree-level
masses, whereas the terms subtracted with minus-signs
are treated as counterterms. For the present problem, however,
resummation is inessential, since it affects only
the contributions coming from the $n=0$-modes.
In other words, it is sufficient to know which
contributions to $V(\varphi)$ come from 3d, but
the exact expressions are not needed.
Hence the thermal corrections to masses may be neglected,
allowing one to treat the temporal and spatial components
of the gauge fields as having the same mass,
which simplifies the expressions somewhat.
Just for cosmetic reasons, one might wish to
calculate the 1-loop contributions from the
thermal counterterms, though, since they cancel the
linear terms of the form $mT^3$ in the unresummed
2-loop $V(\varphi)$, see below.

To calculate $V(\varphi)$, one shifts $\phi\to\phi+\varphi$,
neglects linear terms, and calculates all the
one-particle-irreducible vacuum diagrams.
The masses of the
scalar, vector and fermion fields, respectively,
are of the form
\be
m^2=m_S^2+n_1\lambda\varphi^2,\quad
M^2=n_2 g^2\varphi^2,\quad
m_f^2=n_3g_Y^2\varphi^2, \label{masses}
\ee
where $n_1$, $n_2$, $n_3$ are some numerical factors.
The propagators in eq.~\nr{symprop} change accordingly.
The ghosts remain massless in Landau gauge.
The shift generates mass counterterms ($\delta m^2$,
$\delta M^2$ and $\delta m_f$) from the corresponding
coupling constant counterterms, as well.

To calculate the 1-loop contribution to $V(\varphi)$,
one needs the integrals~\cite{AE}
\ba
J_b(m) & = & \frac{1}{2}\Tint{p_b} \ln ({p^2+m^2})=
\mu^{-2\epsilon}
\biggl[\frac{m^2T^2}{24}-\frac{m^3T}{12\pi}-
\frac{m^4}{64\pi^2}\biggl(\frac{1}{\epsilon}+L_b\biggr)
\biggr]+{\cal O}\Bigl(\frac{m^6}{T^2}\Bigr), \nonumber \\
J_f(m) & = & \frac{1}{2}\Tint{p_f} \ln ({p^2+m^2})=\mu^{-2\epsilon}
\biggl[-\frac{m^2T^2}{48}-
\frac{m^4}{64\pi^2}\biggl(\frac{1}{\epsilon}+L_f\biggr)
\biggr]+{\cal O}\Bigl(\frac{m^6}{T^2}\Bigr). \label{JbJf}
\ea
The terms suppressed by $T^2$
and neglected in eq.~\nr{JbJf} are
\be
J_b^{(6)}=\frac{\zeta(3)}{768\pi^4}\frac{m^6}{T^2},\quad
J_f^{(6)}=\frac{7\zeta(3)}{768 \pi^4}\frac{m^6}{T^2}
\label{hoJ},
\ee
and give
the higher order operators discussed in Sec.~\ref{corrections}.
In the 3d theory, the integral corresponding
to $J_b(m)$ is
\be
J_3(m)=\frac{T}{2}\int \frac{d^dp}{(2\pi)^d}
\ln ({p^2+m^2})=\mu^{-2\epsilon}
\bigg(-\frac{m^3T}{12\pi}\biggr),
\ee
which is just a part of $J_b(m)$. With these integrals, one can write
the typical scalar, vector and fermion contributions
${\cal C}_S(m)$, ${\cal C}_V(M)$ and ${\cal C}_F(m_f)$
to $V_1(\varphi)$, and separate from these the 3d-part.
The massless ghosts do not contribute. The results are
\ba
{\cal C}_S(m) & \equiv &
-\Tint{p_b}\ln \biggl(\frac{1}{p^2+m^2}\biggr)^{1/2}=J_b(m) \nonumber \\
& = & {\cal C}_S^{3d}+
\mu^{-2\epsilon}
\biggl[\frac{m^2T^2}{24}-
\frac{m^4}{64\pi^2}\biggl(\frac{1}{\epsilon}+L_b\biggr)
\biggr],\label{cs} \\
& & \nonumber \\
{\cal C}_V(M) & \equiv &
-\Tint{p_b}\ln \biggl(
\det \frac{\delta_{\mu\nu}-p_\mu p_\nu/p^2}{p^2+M^2}
\biggr)^{1/2}=(3-2\epsilon)J_b(M) \nonumber \\
& = &  {\cal C}_V^{3d}+
\mu^{-2\epsilon}
\biggl[\frac{M^2T^2}{8}-
\frac{M^4}{64\pi^2}\biggl(\frac{3}{\epsilon}+3L_b-2\biggr)
\biggr], \\
& & \nonumber \\
{\cal C}_F(m_f) & \equiv &
\Tint{p_f}\ln \det\frac{1}{i\slash{p}+m_f}=-4 J_f(m_f) \nonumber \\
& = &
\mu^{-2\epsilon}
\biggl[\frac{m_f^2T^2}{12}+
\frac{m_f^4}{16\pi^2}\biggl(\frac{1}{\epsilon}+L_f\biggr)
\biggr],\label{cf}
\ea
where ${\cal C}_S^{3d}$ and ${\cal C}_V^{3d}$
are the corresponding integrals in the 3d theory.
The $1/\epsilon$-parts are $T$-independent, and are
cancelled by the 1-loop counterterms
$\frac{1}{2}\delta m_S^2\varphi^2$, $\frac{1}{4}\delta \lambda\varphi^4$.
Since the bosonic field content of the 3d theory is the same
as that of the original theory, the parts
${\cal C}_S^{3d}$ and ${\cal C}_V^{3d}$
are reproduced by the 3d theory.
The coefficient of $\varphi^2/2$ of the
remaining terms determines the 1-loop
result for $m_3^2$, and the coefficient of $\varphi^4/4$
the 1-loop result for $\lambda_3$.
In this simple way,
the couplings of the fundamental scalar sector
in the effective 3d theory get fixed at 1-loop order.

Next we go to 2-loop level, which is
required for the mass $m_3^2$.
In general, there are three classes of
diagrams (see, e.g., Fig.~23 in~\cite{AE}) contributing
at order $g^4$: the sunset diagrams,
the figure~8 -diagrams, and
the 1-loop counterterm diagrams.
The counterterm diagrams can contain either
the mass or the wave function counterterm.
The general strategy is the same as at 1-loop level: from
each bosonic diagram, one separates the contribution
coming from the $n=0$ modes, since this
contribution is reproduced by the 3d theory.
The remaining part, analytic in the mass parameters,
is not reproduced by the 2-loop diagrams of the 3d theory,
and must hence be due to corrections to the tree-level
parameters of the 3d theory. The fermionic diagrams
do not appear in the 3d theory,
but they are IR-safe, and hence directly produce
terms analytic in the mass parameters, contributing to $m_3^2$.
We will first give the basic integrals appearing
in the calculation, and then the results
for the contributions of the superheavy modes to
all the different types of diagrams that can appear.

The bosonic tadpole integral is
\be
I_b(m)=\Tint{p_b} \frac{1}{p^2+m^2}=I_b'(m)+I_3(m),
\ee
where $I'_b(m)$ is in eq.~\nr{Ib}
and the 3d integral is
\be
I_3(m)=T\int \frac{d^dp}{(2\pi)^d}\frac{1}{p^2+m^2}=\mu^{-2\epsilon}
\bigg(-\frac{mT}{4\pi}\biggr). \label{I3}
\ee
The fermionic tadpole integral
$I_f(m)$ is given in eq.~\nr{If}.
The products appearing in the 2-loop diagrams,
apart from inessential vacuum terms, are
\ba
I_b(m_1)I_b(m_2) & = & I_3(m_1)I_3(m_2)
+\frac{1}{12}\mu^{-2\epsilon}T^2\Bigl[I_3(m_{1})+I_3(m_{2})\Bigr]\nonumber \\
&  & -\mu^{-4\epsilon}\frac{T^2}{16\pi^2}
(m_{1}^2+m_{2}^2)\biggl(
\frac{1}{12\epsilon}+
\frac{L_b}{12}+\frac{\imath_\epsilon}{12}\biggr), \label{f2l} \\
I_b(m)I_f(m_f) & = &
-\frac{1}{24}\mu^{-2\epsilon}T^2 I_3(m)
+ \mu^{-4\epsilon}\frac{T^2}{16\pi^2}\biggl[
m^2\biggl(
\frac{1}{24\epsilon}+
\frac{L_b}{24}+\frac{\imath_\epsilon}{24}-\frac{1}{12}\ln 2\biggr),
\nonumber \\
&  &
-m_f^2\biggl(
\frac{1}{12\epsilon}+
\frac{L_f}{12}+\frac{\imath_\epsilon}{12}\biggr)\biggr], \\
I_f(m_f)I_f(m_{f'}) & = &
\mu^{-4\epsilon}\frac{T^2}{16\pi^2}
(m_f^2+m_{f'}^2)\biggl(
\frac{1}{24\epsilon}+
\frac{L_f}{24}+\frac{\imath_\epsilon}{24}-\frac{1}{12}\ln 2\biggr).
\ea
The bosonic sunset integral is~\cite{FKRS1,AZ}
\ba
H_b(m_1,m_2,m_3) & = &\Tint{p_b,q_b}\frac{1}
{[p^2+m_1^2][q^2+m_2^2][(p+q)^2+m_3^2]} \nonumber \\
& = & \mu^{-4\epsilon}
\frac{T^2}{16\pi^2}\biggl(
\frac{1}{4\epsilon}+\ln\frac{\mu}{m_1+m_2+m_3}+\frac{1}{2}\biggr) \nonumber \\
& = & H_3(m_1,m_2,m_3)+\mu^{-4\epsilon}
\frac{T^2}{16\pi^2}\biggl(
\frac{1}{4}L_b+\frac{1}{4}\imath_\epsilon+
\ln\frac{3T}{\mu}+c\biggr),\label{l2l}
\ea
where $H_3(m_1,m_2,m_3)$ is the corresponding integral
in 3d. Using eq.~\nr{nabi} one can see
that the expression in the brackets
on the last line actually vanishes,
so that $H_3(m_1,m_2,m_3)$ is the second
line in eq.~\nr{l2l}.
However, it proves useful to write
$H_b(m_1,m_2,m_3)$ in the form indicated.
The fermionic integral
\ba
H_f(m_f,m_{f'},m) & = &\Tint{p_f,q_f}\frac{1}
{[p^2+m_f^2][q^2+m_{f'}^2][(p+q)^2+m^2]}
\ea
vanishes~\cite{AE}.

There is also a third integral, $L(m_1,m_2)$, appearing in the
2-loop graphs~\cite{AE}.
Apart from vacuum terms, it is given by
\ba
L(m_1,m_2) & = &
\Tint{p_b,q_b}\frac{(p\cdot q)^2}{p^2(p^2+m_{1}^2)
q^2(q^2+m_{2}^2)}
\nonumber \\
& = & L_3(m_1,m_2)
+\mu^{-2\epsilon}\frac{T^2}{24}\Bigl[I_3(m_{1})+I_3(m_{2})\Bigr]
\nonumber \\
& - & \mu^{-4\epsilon}\frac{T^2}{16\pi^2}
(m_{1}^2+m_{2}^2)
\biggl(
\frac{1}{48\epsilon}+
\frac{L_b}{48}+\frac{\imath_\epsilon}{48}-\frac{1}{48}\biggr),
\ea
where
\be
L_3(m_1,m_2) = \frac{1}{3}I_3(m_1)I_3(m_2).
\ee
However, this integral does not contribute
to the integration over the superheavy scale
in the Standard Model,
since it is cancelled between the figure~8 and sunset
diagrams containing only SU(2)
vector fields (${\cal D}_{\rm VV}$
and ${\cal D}_{\rm VVV}$ below)~\cite{AE}.

Using the given integrals and results from~\cite{AE}
for the 2-loop diagrams, one can write down
the contributions from the superheavy modes to all the
possible types of 2-loop diagrams.
We give here explicit results only for
the simplest mass combinations in the propagators,
relevant for Sec.~\ref{DRinSM};
the results for the cases with other masses
can be read by using eqs.~\nr{f2l}-\nr{l2l} and Appendix A of~\cite{AE}.
As stated above, we need not bother about resummation.
The results are
\ba
{\cal D}_{\rm SSS}(m_{1},m_{2},m_{3}) & = &
\Tint{p_b,q_b}\frac{1}
{[p^2+m_{1}^2][q^2+m_{2}^2][(p+q)^2+m_{3}^2]} \nonumber \\
& = & {\cal D}^{3d}_{\rm SSS}+
\mu^{-4\epsilon}\frac{T^2}{16\pi^2}\biggl(
\frac{L_b}{4}+\frac{\imath_\epsilon}{4}+\ln\frac{3T}{\mu}+c\biggr),
\label{fae} \\
& & \nonumber \\
{\cal D}_{\rm SSV}(m_{1},m_{2},M) & = &
\Tint{p_b,q_b}\frac{(2p_\mu+q_\mu)(2p_\nu+q_\nu)\Bigl(\delta_{\mu\nu}-
\frac{q_\mu q_\nu}{q^2}\Bigr)}
{[p^2+m_{1}^2][q^2+M^2][(p+q)^2+m_{2}^2]}
\nonumber \\
& = & {\cal D}^{3d}_{\rm SSV}+
\frac{1}{6}\mu^{-2\epsilon}T^2I_3(M) \nonumber \\
& &
+\mu^{-4\epsilon}\frac{T^2}{16\pi^2}\biggl[
M^2 \biggl(
-\frac{1}{6\epsilon}+
\frac{L_b}{12}+\frac{\imath_\epsilon}{12}+\ln\frac{3T}{\mu}+c
\biggr) \nonumber \\
&  &
-2(m_{1}^2+m_{2}^2) \biggl(
\frac{L_b}{4}+\frac{\imath_\epsilon}{4}+\ln\frac{3T}{\mu}+c
\biggr)\biggr], \\
& & \nonumber \\
{\cal D}_{\rm SVV}(m,M,M) & = &
\Tint{p_b,q_b}\frac{4\Bigl(\delta_{\mu\nu}-
\frac{p_\mu p_\nu}{p^2}\Bigr)
\Bigl(\delta_{\mu\nu}-\frac{q_\mu q_\nu}{q^2}\Bigr)}
{[p^2+M^2][q^2+M^2][(p+q)^2+m^2]}
\nonumber \\
& = & {\cal D}^{3d}_{\rm SVV}+
\mu^{-4\epsilon}\frac{T^2}{16\pi^2}\biggl[
\frac{5}{2}L_b+\frac{5}{2}\imath_\epsilon+
10\biggl(\ln\frac{3T}{\mu}+c\biggr)
\biggr], \\
& & \nonumber \\
{\cal D}_{\rm VVV}(M,M,M) & = & \!\!\!\!
\Tint{p_b,q_b}\!\!\!\frac{
\Bigl(\delta_{\mu\alpha}-\frac{p_\mu p_{\alpha}}{p^2}\Bigr)
\Bigl(\delta_{\nu\beta}-\frac{q_\nu q_{\beta}}{q^2}\Bigr)
\Bigl(\delta_{\rho\gamma}-\frac{r_\rho r_{\gamma}}{r^2}\Bigr)}
{[p^2+M^2][q^2+M^2][(p+q)^2+M^2]}
F_{\mu\nu\rho}(p,q,r)F_{\alpha\beta\gamma}(p,q,r)
\nonumber \\
& = & \tilde{\cal D}^{3d}_{\rm VVV}+\frac{7}{4}\mu^{-2\epsilon}T^2I_3(M)
-3L(M,M) \nonumber \\
&  &
-\mu^{-4\epsilon}\frac{T^2}{16\pi^2}M^2 \biggl[
\frac{7}{4}\frac{1}{\epsilon}+
\frac{31}{4}L_b+\frac{31}{4}\imath_\epsilon+
24\biggl(\ln\frac{3T}{\mu}+c\biggr)-1\biggr], \\
& & \nonumber \\
{\cal D}_{\rm \eta\eta V}(M) & = &-
\Tint{p_b,q_b}\frac{p_\mu(p_\nu+q_\nu)\Bigl(\delta_{\mu\nu}-
\frac{q_\mu q_\nu}{q^2}\Bigr)}
{p^2(q^2+M^2)(p+q)^2}
\nonumber \\
& = & {\cal D}^{3d}_{\rm \eta\eta V}-
\frac{1}{24}\mu^{-2\epsilon}T^2 I_3(M) \nonumber \\
&  &
-\mu^{-4\epsilon}\frac{T^2}{16\pi^2}M^2\biggl[
-\frac{1}{24\epsilon}+
\frac{L_b}{48}+\frac{\imath_\epsilon}{48}+
\frac{1}{4}\biggl(\ln\frac{3T}{\mu}+c\biggr)\biggr], \\
& & \nonumber \\
{\cal D}_{\rm FFS}(m_{f},m_{f'},m) & = &
\Tint{p_f,q_f}\frac{\mathop{\rm Tr}(i\slash{p}+m_{f})
(i\slash{q}-m_{f'})}
{[p^2+m_{f}^2][q^2+m_{f'}^2][(p+q)^2+m^2]}
\nonumber \\
& = & -\frac{1}{6}\mu^{-2\epsilon}T^2I_3(m)+
\mu^{-4\epsilon}\frac{T^2}{16\pi^2}\biggl[
m^2\biggl(\frac{1}{6\epsilon}+
\frac{L_b}{6}+\frac{\imath_\epsilon}{6}-\frac{1}{3}\ln 2\biggr)
\nonumber \\
&  &
-(m_{f}^2+m_{f'}^2)\biggl(\frac{1}{4\epsilon}+
\frac{L_f}{4}+\frac{\imath_\epsilon}{4}-\frac{1}{6}\ln 2\biggr)
\biggr], \\
& & \nonumber \\
{\cal D}_{\rm FFV}(m_{f},m_{f'},M) & = &
-\Tint{p_f,q_f}\frac{\mathop{\rm Tr}(i\slash{p}+m_{f})
\gamma_\mu a_L(i\slash{q}-m_{f'})\gamma_\nu a_L}
{[p^2+m_{f}^2][q^2+m_{f'}^2][(p+q)^2+M^2]}
\biggl(\delta_{\mu\nu}-\frac{
r_\mu r_\nu}{r^2}\biggr)
\nonumber \\
& = &
-\frac{1}{6}\mu^{-2\epsilon}T^2I_3(M)+
\mu^{-4\epsilon}\frac{T^2}{16\pi^2}\biggl[
M^2\biggl(\frac{1}{6\epsilon}+
\frac{L_b}{6}+\frac{\imath_\epsilon}{6}-\frac{1}{3}\ln 2-\frac{1}{6}\biggr)
\nonumber \\
&  &
-(m_{f}^2+m_{f'}^2)\biggl(\frac{1}{4\epsilon}+
\frac{L_f}{4}+\frac{\imath_\epsilon}{4}-\frac{1}{6}\ln 2-\frac{1}{4}\biggr)
\biggr], \\
& & \nonumber \\
{\cal D}_{\rm SS}(m_{1},m_{2})  & = &
-\Tint{p_b,q_b}\frac{1}{(p^2+m_{1}^2)(q^2+m_{2}^2)}
\nonumber \\
& = & {\cal D}^{3d}_{\rm SS}
-\frac{1}{12}\mu^{-2\epsilon}T^2\Bigl[I_3(m_{1})+I_3(m_{2})\Bigr]\nonumber \\
&  & +\mu^{-4\epsilon}\frac{T^2}{16\pi^2}
(m_{1}^2+m_{2}^2)
\biggl(
\frac{1}{12\epsilon}+
\frac{L_b}{12}+\frac{\imath_\epsilon}{12}\biggr), \\
& & \nonumber \\
{\cal D}_{\rm SV}(m,M)  & = &
-2\;\Tint{p_b,q_b}\frac{\delta_{\mu\nu}
\Bigl(\delta_{\mu\nu}-\frac{q_\mu q_\nu}{q^2}\Bigr)}
{(p^2+m^2)(q^2+M^2)}
\nonumber \\
& = & {\cal D}^{3d}_{\rm SV}
-\frac{1}{2}\mu^{-2\epsilon}T^2\Bigl[I_3(m)+I_3(M)\Bigr] \nonumber \\
&  & +\mu^{-4\epsilon}\frac{T^2}{16\pi^2}
(m^2+M^2)
\biggl(
\frac{1}{2\epsilon}+
\frac{L_b}{2}+\frac{\imath_\epsilon}{2}-\frac{1}{3}\biggr), \\
& & \nonumber \\
{\cal D}_{\rm VV}(M_{1},M_{2})  & = &
-\Tint{p_b,q_b}\frac{\Bigl(\delta_{\mu\nu}-\frac{p_\mu p_\nu}{p^2}\Bigr)
\Bigl(\delta_{\sigma\rho}-\frac{q_\sigma q_\rho}{q^2}\Bigr)}
{(p^2+M_{1}^2)(q^2+M_{2}^2)}
(2\delta_{\mu\nu}\delta_{\sigma\rho}-
\delta_{\mu\sigma}\delta_{\nu\rho}-
\delta_{\mu\rho}\delta_{\nu\sigma})
\nonumber \\
& = & \tilde{\cal D}^{3d}_{\rm VV}
-\frac{7}{6}\mu^{-2\epsilon}T^2\Bigl[I_3(M_{1})+I_3(M_{2})\Bigr]+
2L(M_{1},M_{2}) \nonumber \\
&  & +\mu^{-4\epsilon}\frac{T^2}{16\pi^2}
(M_{1}^2+M_{2}^2)
\biggl(
\frac{7}{6\epsilon}+
\frac{7}{6}L_b+\frac{7}{6}\imath_\epsilon-\frac{5}{3}\biggr), \\
& & \nonumber \\
{\cal D}_S(m) & = &
\Tint{p_b}\frac{\delta m^2+\delta Z_S p^2}{p^2+m^2}
=\mu^{-2\epsilon}\frac{T^2}{12}(1+\epsilon \imath_\epsilon)
(\delta m^2-m^2\delta Z_S),\\
& & \nonumber \\
{\cal D}_V(M) & = &
\Tint{p_b}\frac{\delta M^2+\delta Z_V p^2}{p^2+M^2}
\delta_{\mu\nu}\Bigl(\delta_{\mu\nu}-\frac{p_\mu p_\nu}{p^2}\Bigr) \nonumber \\
& = & \mu^{-2\epsilon}\frac{T^2}{12}
[3+3\epsilon\imath_\epsilon-2\epsilon]
(\delta M^2-M^2\delta Z_V),\\
& & \nonumber \\
{\cal D}_F(m_f) & = &
\Tint{p_f}\frac{1}{p^2+m_f^2}
\mathop{\rm Tr} [i\slash{p}-m_f]
[\delta m_f+i\slash{p}(a_L\delta Z^L_F+a_R\delta Z^R_F)] \nonumber \\
& = & \mu^{-2\epsilon}\frac{T^2}{12}
[1+\epsilon (\imath_\epsilon-2\ln 2)]
[\delta m_f^2-m_f^2(\delta Z^L_F+\delta Z^R_F)]. \label{lae}
\ea
In the expressions for
${\cal D}_{\rm VVV}$ and ${\cal D}_{\rm FFV}$,
we used $r$ as a shorthand for $-p-q$.
The tilde over $\tilde{\cal D}^{3d}_{\rm VVV}$ and
$\tilde{\cal D}^{3d}_{\rm VV}$ indicates
that the 3d-part of $L(M,M)$ is not included.
In eq.~\nr{lae}, $\delta m_f^2 \equiv 2 m_f \delta m_f$.

In addition to the diagrams mentioned above,
one may calculate the thermal counterterm diagrams.
According to eq.~\nr{TCT}, they give
contributions of the form
\be
-\overline{\Pi}_{\phi}(0)I_3(m),\quad
-\overline{\Pi}_{A_0}(0)I_3(M). \label{thctc}
\ee
These contributions cancel the linear terms proportional
to $T^2I_3(m)$, $T^2I_3(M)$
in eqs.~\nr{fae}-\nr{lae}.
After this cancellation, all that is left
is terms proportional to the masses squared.
Using eq.~\nr{masses},
such terms give the coefficient of $\varphi^2/2$, i.e.,
the mass $m_3^2$ of the 3d theory.
The counterterm contributions
${\cal D}_S(m)$, ${\cal D}_V(M)$, ${\cal D}_F(m_f)$
do not cancel all the $1/\epsilon$-parts from the
other contributions, since there remains a 2-loop mass
counterterm in the 3d theory~\cite{FKRS1}.

\subsubsection{The couplings of the adjoint scalar sector}
\label{adjoint}

The couplings of the adjoint scalar sector
could be calculated from an effective potential
as for the fundamental scalar sector, but
for completeness we calculate them directly from Feynman diagrams.
We make here just a 1-loop calculation.

\begin{figure}[tbh]

\hspace*{1.0cm}
\epsfysize=7cm
\epsffile[100 500 500 700]{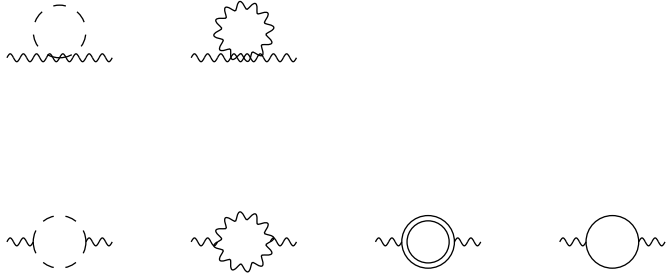}
\vspace*{-1cm}

\caption[a]{\protect
The diagrams needed for calculating
the 3d mass
of the temporal components
of the gauge fields.}
\label{drmd}
\end{figure}

The 1-loop diagrams needed for calculating
the $(A_0A_0)$-correlator
at vanishing external momenta are shown in Fig.~\ref{drmd}.
We need two new integrals, obtained
by taking the derivative of eqs.~\nr{Ib}, \nr{If}
with respect to~$T$. With the accuracy needed,
\ba
E_b'(m) & = & \Tint{p_b}'\frac{p_0^2}{(p^2+m^2)^2}=
-\frac{T^2}{24}-
\frac{m^2}{16\pi^2}\biggl(\frac{1}{2\epsilon_b}+1\biggr)
,\nonumber \\
E_f(m) & = &\Tint{p_f}\frac{p_0^2}{(p^2+m^2)^2}=
\frac{T^2}{48}-
\frac{m^2}{16\pi^2}\biggl(\frac{1}{2\epsilon_f}+1\biggr).
\label{Es}
\ea
Note that $E_b'(0)$ is the constant part
subtracted in the definition of $J^b_{00}$
in eq.~\nr{integrals}.
Using the integrals in eq.~\nr{Es} together with those in
eqs.~\nr{Ib},~\nr{integrals}, the diagrams in Fig.~\ref{drmd} give
\ba
{\cal A}^{(2)}_{\rm S} & = &
\Tint{p_b}'\frac{1}{p^2+m_S^2} = I_b'(m_S) =\frac{T^2}{12}
-\frac{m_S^2}{16\pi^2}\frac{1}{\epsilon_b}, \label{a2s}
\\ & & \nonumber \\
{\cal A}^{(2)}_{\rm V} & = &
\Tint{p_b}'\frac{\delta_{\mu\nu}-\frac{p_\mu p_\nu}{p^2}}{p^2}
\Bigl(2\delta_{00}\delta_{\mu\nu}-2\delta_{0\mu}\delta_{0\nu}\Bigr)
= 4(1-\epsilon)I_b'(0)+2E_b'(0) =\frac{T^2}{4},
\\ & & \nonumber \\
{\cal A}^{(2)}_{\rm SS} & = &
\Tint{p_b}'\frac{(2p_0)^2}{(p^2+m_S^2)^2}
= 4E_b'(m_S) =-\frac{T^2}{6}-\frac{m_S^2}{16\pi^2}
\biggl(\frac{2}{\epsilon_b}+4\biggr),
\\ & & \nonumber \\
{\cal A}^{(2)}_{\rm VV} & = &
\Tint{p_b}'\frac{(12-8\epsilon)p_0^2}{(p^2)^2}
= (12-8\epsilon)E_b'(0) =-\frac{T^2}{2},
\\ & & \nonumber \\
{\cal A}^{(2)}_{\eta\eta} & = &
\Tint{p_b}'\frac{p_0^2}{(p^2)^2}
= E_b'(0) =-\frac{T^2}{24},
\\ & & \nonumber \\
{\cal A}^{(2)}_{\rm FF} & = &
-2\Tint{p_f}\frac{2p_0^2-p^2}{(p^2)^2}
= -4E_f(0)+2I_f(0) =-\frac{T^2}{6}. \label{a2ff}
\ea
For ${\cal A}^{(2)}_{\rm SS}$,
${\cal A}^{(2)}_{\rm VV}$,
${\cal A}^{(2)}_{\rm \eta\eta}$, and
${\cal A}^{(2)}_{\rm FF}$,
the integrand was obtained from eqs.~\nr{za0ss}, \nr{Muv},
\nr{za0ee}, and \nr{za0ff} by putting $k\to 0$.
In the final result, the $1/\epsilon$-parts cancel
between ${\cal A}^{(2)}_{\rm S}$ and ${\cal A}^{(2)}_{\rm SS}$.
The constant term proportional to $m_S^2$
is of higher order, and the $T^2$-terms
give the desired 3d mass parameter.

\begin{figure}[tbh]

\hspace*{2.0cm}
\epsfysize=8.5cm
\epsffile[100 440 500 700]{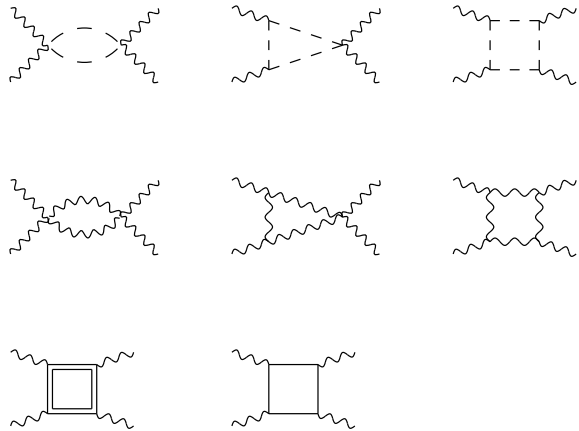}
\vspace*{-0.5cm}

\caption[a]{\protect
The diagrams needed for calculating
the quartic self-coupling
of the temporal components
of the gauge fields.}
\label{drla}
\end{figure}

The diagrams needed for the
$(A_0A_0A_0A_0)$-correlator are shown in Fig.~\ref{drla}.
They give
\ba
{\cal A}^{(4)}_{\rm SS} & = &
\Tint{p_b}'
\frac{1}{(p^2)^2}
=B_b'=
\frac{1}{16\pi^2}
\frac{1}{\epsilon_b}, \label{a4ss}
\\ & & \nonumber \\
{\cal A}^{(4)}_{\rm SSS} & = &
\Tint{p_b}'
\frac{(2p_0)^2}{(p^2)^3}
=4K^b_{00}=
\frac{1}{16\pi^2}
\biggl(\frac{1}{\epsilon_b}+2\biggr),
\\ & & \nonumber \\
{\cal A}^{(4)}_{\rm SSSS} & = &
\Tint{p_b}'
\frac{(2p_0)^4}{(p^2)^4}
=16L^b_0=
\frac{1}{16\pi^2}
\biggl(\frac{2}{\epsilon_b}+\frac{16}{3}\biggr),
\\ & & \nonumber \\
{\cal A}^{(4)}_{\rm VV} & = &
\Tint{p_b}'
\frac{
\Bigl(\delta_{\mu\nu}-\frac{p_\mu p_\nu}{p^2}\Bigr)
\Bigl(\delta_{\sigma\rho}-\frac{p_\sigma p_\rho}{p^2}\Bigr)}{(p^2)^2}
\nonumber \\
& &
\nonumber \\
& & \times
\Bigl(2\delta_{00}\delta_{\mu\sigma}-
2\delta_{0\mu}\delta_{0\sigma}\Bigr)
\Bigl(2\delta_{00}\delta_{\nu\rho}-
2\delta_{0\nu}\delta_{0\rho}\Bigr)
\nonumber \\
& = &
8 (1-\epsilon) B_b'+
4L^b_0=
\frac{1}{16\pi^2}
\biggl(\frac{17}{2}\frac{1}{\epsilon_b}-\frac{20}{3}\biggr),
\\ & & \nonumber \\
{\cal A}^{(4)}_{\rm VVV} & = &
\Tint{p_b}'
\frac{
\Bigl(\delta_{\mu\nu}-\frac{p_\mu p_\nu}{p^2}\Bigr)
\Bigl(\delta_{\sigma\alpha}-\frac{p_\sigma p_\alpha}{p^2}\Bigr)
\Bigl(\delta_{\rho\beta}-\frac{p_\rho p_\beta}{p^2}\Bigr)}{(p^2)^3}
\nonumber \\
& &
\nonumber \\
& & \times
F_{0\mu\sigma}(0,p,-p)
F_{0\nu\rho}(0,p,-p)
\Bigl(2\delta_{00}\delta_{\alpha\beta}-
2\delta_{0\alpha}\delta_{0\beta}\Bigr)
\nonumber \\
& = &
16 (1-\epsilon) K^b_{00}+
8L^b_0=
\frac{1}{16\pi^2}
\biggl(\frac{5}{\epsilon_b}+\frac{20}{3}\biggr),
\\ & & \nonumber \\
{\cal A}^{(4)}_{\rm VVVV} & = &
\Tint{p_b}'
\frac{
\Bigl(\delta_{\alpha\beta}-\frac{p_\alpha p_\beta}{p^2}\Bigr)
\Bigl(\delta_{\gamma\delta}-\frac{p_\gamma p_\delta}{p^2}\Bigr)
\Bigl(\delta_{\mu\nu}-\frac{p_\mu p_\nu}{p^2}\Bigr)
\Bigl(\delta_{\sigma\rho}-\frac{p_\sigma p_\rho}{p^2}\Bigr)}{(p^2)^4}
\nonumber \\
& &
\nonumber \\
& & \times
F_{0\alpha\gamma}(0,p,-p)
F_{0\delta\nu}(0,p,-p)
F_{0\mu\sigma}(0,p,-p)
F_{0\beta\rho}(0,p,-p) \nonumber \\
& = &
16 (3-2\epsilon) L^b_0=
\frac{1}{16\pi^2}
\biggl(\frac{6}{\epsilon_b}+12\biggr),
\\ & & \nonumber \\
{\cal A}^{(4)}_{\rm \eta\eta\eta\eta} & = &
\Tint{p_b}'
\frac{p_0^4}{(p^2)^4}
=L^b_0=
\frac{1}{16\pi^2}
\biggl(\frac{1}{8\epsilon_b}+\frac{1}{3}\biggr),
\\ & & \nonumber \\
{\cal A}^{(4)}_{\rm FFFF} & = &
\Tint{p_f}\frac{1}{(p^2)^4}
\mathop{\rm Tr} [(i\slash{p}\gamma_0 a_L)(i\slash{p}\gamma_0 a_L)
(i\slash{p}\gamma_0 a_L)(i\slash{p}\gamma_0 a_L)] \nonumber \\
& = &
=16L^f_0-16K^f_{00}+2B_f'
=\frac{1}{16\pi^2}
\biggl(-\frac{8}{3}\biggr). \label{a4ffff}
\ea
Again the $1/\epsilon$-parts cancel
when the correct coefficients are included,
since there are no coupling constant
counterterms in the 3d theory.

\subsection{Integration over the heavy scale}

The integration over the heavy scale proceeds analogously
to the integration over the superheavy scale.
Instead of an infinite number of excitations
with masses $\sim \pi T$, there are
now a finite number of fields with
masses $m_D\sim gT$. The heavy fields include, in particular,
the temporal components $A_0$ of the gauge fields.
The heavy fields are scalars
in the effective 3d theory, so
no gauge fixing is needed for the integration.
The general form of the resulting theory
differs from the starting point
only by the absence of the heavy fields,
so that the final theory is the light bosonic
sector of the original 4d theory,
but in three spatial dimensions.

There are three kinds of vertices which
the heavy field $A_0$ feels.
The interactions with the spatial gauge fields
follow from eq.~\nr{gaugevs}, and are of the form
\be
ig_3f^{abc}A_0^a(p)A_i^b(q)A_0^c(r)(p_i-r_i),\quad
g_3^2 G^{abcd}_{ij00}A^a_i A^b_j A^c_0A^d_0,
\label{a0ai}
\ee
where $G^{abcd}_{ij00}\propto\delta_{ij}$.
Here $g_3$ denotes the dimensionful
3d gauge coupling, and the fields have also
been scaled by $T$ to have the dimension GeV$^{1/2}$.
In addition to eq.~\nr{a0ai} there are different
quartic scalar vertices. One of the scalar vertices,
the quartic self-interaction of $A_0$, can be neglected,
since the coupling constant is of order $g^4$,
and would appear only inside 2-loop graphs
where it is further suppressed by other couplings.

The integration measure in 3d is denoted by
\be
\int dp \equiv \int\frac{d^dp}{(2\pi)^d}.
\ee
The $A_0$-propagator in the symmetric phase is
\be
\langle A_0(p)A_0(-p)\rangle=
\frac{1}{p^2+m_D^2}.
\ee
The basic integrals needed are $\bar{I}_3(m)$, which
is just eq.~\nr{I3} without $T$, together with
${\cal B}(k^2;m_1,m_2)$ and ${\cal J}_{ij}$, defined by
\ba
{\cal B}(k^2;m_1,m_2) & = &
\int dp\frac{1}{[p^2+m_1^2][(p+k)^2+m_2^2]}\nonumber \\
& = &
\frac{i}{8\pi (k^2)^{1/2}}
\ln\frac{m_1+m_2-i(k^2)^{1/2}}{m_1+m_2+i(k^2)^{1/2}}
\nonumber \\
& = &
\frac{1}{4\pi(m_1+m_2)}\biggl[
1-\frac{1}{3}\frac{k^2}{(m_1+m_2)^2}+\ldots
\biggr],
\\ && \nonumber \\
{\cal J}_{ij} & = &
\int dp\frac{p_ip_j}{[p^2+m_D^2][(p+k)^2+m_D^2]}-
\int dp\frac{p_ip_j}{(p^2+m_D^2)^2}\nonumber \\
& = &
-\biggl(\delta_{ij}-\frac{k_ik_j}{k^2}\biggr)
\frac{k^2}{96\pi m_D}+
\frac{k_ik_j}{32 \pi m_D}+{\cal O}\Bigl(\frac{k^4}{m_D^3}\Bigr).
\ea
Using these integrals, the task is to calculate
the corrections from $m_D$ to the parameters of
the final theory in powers of $g_3^2/m_D$, where
it is assumed that the light masses and momenta
are $m\sim k\sim g_3^2$.

Let us start with wave function normalization.
Since the fundamental scalar field interacts
with $A_0$ only through quartic vertices,
there is no momentum-dependent correction from
the $A_0$-field to the $\phi$-correlator.
Hence the normalization of $\phi$ does not change.
The momentum-dependent correction to the $A_i$-correlator
comes from the diagram with two internal $A_0$-propagators,
and is, in analogy with ${\cal Z}^{A_i}_{\rm SS}$
in eq.~\nr{zaiss},
\ba
{\cal Z}^{A_i}_{\rm LL}  & = &
\int dp\frac{(2p_i+k_i)(2p_j+k_j)}
{[p^2+m_D^2][(p+k)^2+m_D^2]}\nonumber \\
& \Rightarrow & 4{\cal J}^{(T)}_{ij}
=-\biggl(\delta_{ij}-\frac{k_ik_j}{k^2}\biggr)
\frac{k^2}{24\pi m_D}. \label{zaill}
\ea
The heavy propagators are denoted by L.

To fix the gauge coupling, one can calculate
the contributions of $A_0$ to the $(\phi\phi A_iA_j)$-vertex.
There are two types of diagrams, in analogy with
${\cal G}^{A_i}_{\rm SS}$ and ${\cal G}^{A_i}_{\rm SSS}$
in eqs.~\nr{gaiss}, \nr{gaisss}.
The contributions are
\ba
{\cal G}_{\rm LL} & = & \int dp \frac{\delta_{ij}}{(p^2+m_D^2)^2}=
\frac{\delta_{ij}}{8\pi m_D}, \label{gll} \\
{\cal G}_{\rm LLL} & = &
\int dp \frac{(2p_i)(2p_j)}{(p^2+m_D^2)^3}=
\frac{\delta_{ij}}{8\pi m_D}. \label{glll}
\ea

It is very easy to calculate the 1-loop
corrections to the parameters of the scalar sector,
since the field one is integrating over is
itself a scalar. Hence the diagrams can give
just $\bar{I}_3(m_D)$ or ${\cal B}(0;m_D,m_D)$.
For the mass parameter, one needs again a 2-loop
calculation, and it is best to employ
the effective potential. After the shift,
the mass of the heavy field is of
the form $m_L^2=m_D^2+h_3\varphi^2$,
where $h_3$ is the coupling of the $(\phi\phi A_0A_0)$-interaction.
The diagrams with heavy fields in internal lines
do not exist in the final theory, so their effect has to be
produced by different parameters in the action.
One needs to expand the results of
these diagrams in powers of $m/m_D$ to such order that
terms of the form $m_D m$ and $m^2$ are kept.
Constant terms proportional to $m_D^2$ ar neglected.
For the expansion, one writes
\be
m_L=m_D+\frac{1}{2}\frac{h_3\varphi^2}{m_D}+\ldots\, .
\ee
There are four types of possible 2-loop
diagrams, and they give~\cite{FKRS1}
\ba
D_{\rm LS}(m) & = &
\int dp\,dq\frac{1}{(p^2+m_L^2)(q^2+m^2)} \nonumber \\
& = & \bar{I}_3(m_L)\bar{I}_3(m)\Rightarrow
-\frac{\mu^{-2\epsilon}}{4 \pi} m_D \bar{I}_3(m), \label{dls}
\\ & & \nonumber \\
D_{\rm LV}(M) & = &
\int dp\,dq\frac{\delta_{ij}
\Bigl(\delta_{ij}-\frac{q_iq_j}{q^2}\Bigr)}
{(p^2+m_L^2)(q^2+M^2)} \nonumber \\
& = & (2-2\epsilon)\bar{I}_3(m_L)\bar{I}_3(M)\Rightarrow
-\frac{\mu^{-2\epsilon}}{2 \pi} m_D \bar{I}_3(M),
\\ & & \nonumber \\
D_{\rm LLS}(m) & = &
\int dp\,dq\frac{1}
{[p^2+m_L^2][q^2+m_L^2][(p+q)^2+m^2]} \nonumber \\
& = & \bar{H}_3(m_L,m_L,m)\Rightarrow
\frac{\mu^{-4\epsilon}}{16 \pi^2}
\biggl(\frac{1}{4\epsilon}+
\ln\frac{\mu}{2 m_D}+\frac{1}{2}\biggr),
\\ & & \nonumber \\
D_{\rm LLV}(M) & = &
\int dp\,dq\frac{(2p_i+q_i)(2p_j+q_j)
\Bigl(\delta_{ij}-\frac{q_iq_j}{q^2}
\Bigr)}
{[p^2+m_L^2][q^2+M^2][(p+q)^2+m_L^2]} \nonumber \\
& = & (M^2-4m_L^2)\bar{H}_3(m_L,m_L,M)+
2\bar{I}_3(m_L)\bar{I}_3(M)-\bar{I}_3(m_L)\bar{I}_3(m_L) \nonumber \\
& \Rightarrow &
-\frac{\mu^{-2\epsilon}}{\pi} m_D \bar{I}_3(M)\nonumber \\
& & +
\frac{\mu^{-4\epsilon}}{16 \pi^2}
\biggl[
M^2\biggl(\frac{1}{4\epsilon}+
\ln\frac{\mu}{2 m_D}\biggr)-
h_3\varphi^2\biggl(\frac{1}{\epsilon}+
4\ln\frac{\mu}{2 m_D}+1\biggr) \label{dllv}
\biggr].
\ea
Here $\bar{H}_3$ is the $H_3$ of eq.~\nr{l2l}
divided by $T^2$.
The  ``linear'' terms proportional to $m_D \bar{I}_3(m)$ in
$D_{\rm LV}$ and $D_{\rm LLV}$ cancel each other, and those in
$D_{\rm LS}$ are cancelled by corrections from $m_D$
to the mass of the scalar field
inside 1-loop diagrams.
This is in complete analogy with the cancellation
of linear terms in the integration over
the superheavy scale by the thermal counterterms.
The actual effect then comes from the diagrams
$D_{\rm LLS}$ and $D_{\rm LLV}$. The $1/\epsilon$-parts
account for the change in the mass counter\-term,
and the rest gives the change
in the renormalized mass parameter.

\section{Dimensional reduction in the Standard Model}
\label{DRinSM}

In this Section we add the correct coefficients,
relevant for the Standard Model,
to the generic results of Sec.~\ref{blocks}.
The coefficients are composed of
combinatorial factors, group theoretic factors from
isospin contractions,
and of coupling constants.
We also explain in detail how the
final 3d theory is constructed.

\subsection{Notation}

We treat the Standard Model
with the Higgs sector
\be
{\cal L}_s=(D_{\mu}\Phi)^{\dagger}D_\mu\Phi-\nu^2\Phi^\dagger\Phi+
\lambda\Bigl(\Phi^\dagger\Phi\Bigr)^2
\ee
in the following consistent approximation.
We take $g_Y\neq 0$ only for the top quark, so
that the Yukawa sector is
\be
{\cal L}_Y= g_Y \bigl( \bar{q}_L \tilde{\Phi} t_R +
\bar{t}_R \tilde{\Phi}^{\dagger} q_L \bigr).
\ee
Here $\tilde{\Phi}=i\tau_2\Phi^*$, $\tau_2$ is a Pauli matrix,
and $\Phi$ is the Higgs doublet.
The gauge couplings are denoted by~$g$,
$g'$, and $g_S$. We use the formal
power-counting rule $g'^2\sim g^3$, and keep contributions
only below order $g^5$. This allows one to neglect $g'^2$
e.g. inside 1-loop formulas
of vacuum renormalization, so that $m_W=m_Z$ there.

Most of the counterterms of the Standard Model
can be read from eq.~(A11) of~\cite{AE}.
For completeness, let us state the bare parameters of the Higgs sector,
since the terms proportional to $\lambda$ were neglected in~\cite{AE}:
\ba
\lambda_B & = & \lambda+\frac{\mu^{-2\epsilon}}{(4\pi )^2\epsilon}
\biggl(\frac{9}{16}g^4-\frac{9}{2}\lambda g^2+12\lambda^2
-3g_Y^4+6\lambda g_Y^2\biggr), \\
{\nu}^2_B & = & \nu^2\biggl[1-
\frac{\mu^{-2\epsilon}}{(4\pi )^2\epsilon}\biggl(
\frac{9}{4}g^2-6\lambda-3g_Y^2\biggr)\biggr].
\ea
Here the coupling constants have not been scaled to be dimensionless
in contrast to~\cite{AE}. Note that our
convention for $\lambda$ differs from~\cite{AE}
additionally by the factor~6.
Let us also write down a few useful combinations of the bare
parameters:
\ba
\nu^2_B\phi_B^2 & = & \nu^2\phi^2
\biggl[1+ \frac{\mu^{-2\epsilon}}{(4\pi )^2\epsilon}
6\lambda
\biggr], \label{dnf} \\
\lambda_B\phi_B^4 & = & \phi^4
\biggl[\lambda+ \frac{\mu^{-2\epsilon}}{(4\pi )^2\epsilon}
\biggl(
\frac{9}{16}g^4+12\lambda^2-3g_Y^4
\biggr)
\biggr], \\
g_B^2\phi_B^2A_B^2 & = & g^2\phi^2A^2
\biggl[1- \frac{\mu^{-2\epsilon}}{(4\pi )^2\epsilon}
\biggl(
\frac{3}{4}g^2+3g_Y^2
\biggr)
\biggr], \\
g_{Y,B}\bar{q}_{L,B}\tilde{\Phi}_Bt_{R,B} & = & g_Y\bar{q}_L
\tilde{\Phi} t_R
\biggl[1- \frac{\mu^{-2\epsilon}}{(4\pi )^2\epsilon}
4 g_S^2
\biggr]. \label{dgY}
\ea

Within our approximation, the Standard
Model contains five running parameters,
$\nu^2(\mu)$, $g^2(\mu)$, $\lambda (\mu)$, $g_Y^2(\mu)$,
and $g_S^2(\mu)$.
The first four run at 1-loop order as
\ba
\mu\frac{d}{d\mu}\nu^2(\mu) & = & \frac{1}{8\pi^2}
\biggl(-\frac{9}{4}g^2+6\lambda+3g_Y^2\biggr)\nu^2, \label{run1} \\
\mu\frac{d}{d\mu}g^2(\mu) & = & \frac{1}{8\pi^2}
\biggl(\frac{8n_F+N_s-44}{6}\biggr)g^4, \\
\mu\frac{d}{d\mu}\lambda(\mu) & = & \frac{1}{8\pi^2}
\biggl(\frac{9}{16}g^4-\frac{9}{2}\lambda g^2+12\lambda^2-
3g_Y^4+6\lambda g_Y^2\biggr), \\
\mu\frac{d}{d\mu}g_Y^2(\mu) & = & \frac{1}{8\pi^2}
\biggl(\frac{9}{2}g_Y^4-\frac{9}{4}g^2g_Y^2-8g_S^2g_Y^2\biggr).
\label{run4}
\ea
Here $n_F=3$ is the number of families
and $N_s=1$ is the number of Higgs doublets.
The running of the strong
coupling $g_S$ is not explicitly needed for the
present problem, since $g_S$ appears only inside
loop corrections.
The running of $g'^2$ is of higher order
according to our convention.
The fields $\Phi$, $A$ and $\psi$ run as well, and
the formulas can be extracted, e.g.,  from
eqs.~\nr{dnf}-\nr{dgY},~\nr{run1}-\nr{run4}.

To simplify some of the formulas below, we will
use extensively the notation
\be
h\equiv \frac{m_H}{m_W},\quad
t\equiv \frac{m_t}{m_W}, \quad
s\equiv \frac{g_S}{g}, \label{hts}
\ee
where $m_W$, $m_H$ and $m_t$ are the physical
W boson, Higgs particle and top quark masses.
Inside 1-loop formulas one may use the
tree level relations $g^2h^2=8\lambda$ and $g^2t^2=2g_Y^2$.

\subsection{Integration over the superheavy scale}
\label{ioss}

For the SU(2)+Higgs model, the formulas for dimensional
reduction to order~$g^4$ have been given in~\cite{FKRS1}.
We add here the effect of fermions, and correct an error
coming from~\cite{L}.

Due to 3d gauge invariance, the
effective 3d theory is of the form
\ba
S & = & \int\! d^3x \biggl\{
\frac{1}{4}G^a_{ij}G^a_{ij}+ \frac{1}{4}F_{ij}F_{ij}+
(D_i\Phi)^{\dagger}(D_i\Phi)+
m_3^2\Phi^{\dagger}\Phi+\lambda_3
(\Phi^{\dagger}\Phi)^2 \nonumber \\
& &\hspace*{1.0cm} +\frac{1}{2} (D_iA_0^a)^2+\frac{1}{2}m_D^2A_0^aA_0^a+
\frac{1}{4}\lambda_A(A_0^aA_0^a)^2
+\frac{1}{2} (\partial_iB_0)^2+\frac{1}{2}m_D'^2B_0B_0 \nonumber\\
& &\hspace*{1.0cm}+ h_3\Phi^{\dagger}\Phi A_0^aA_0^a
+ h_3'\Phi^{\dagger}\Phi B_0B_0
-\frac{1}{2}g_3g_3'B_0 \Phi^{\dagger}A_0^a\tau^a\Phi
\,\,\biggr\} ,
\label{action}
\ea
where $G^a_{ij}=\partial_iA_j^a-\partial_jA_i^a+g_3\epsilon^{abc}A^b_iA^c_j$,
$F_{ij}=\partial_iB_j-\partial_jB_i$,
$D_i\Phi=(\partial_i-ig_3\tau^aA^a_i/2+ig_3'B_i/2)\Phi$,
$D_iA_0^a=\partial_iA_0^a+g_3\epsilon^{abc}A_i^bA_0^c$, and
$\Phi=(\phi_3+i\phi_4,\phi_1+i\phi_2)^T/\sqrt{2}$.
The $\tau^a$:s are the Pauli matrices.
The factor $1/T$ multiplying the action has been scaled into
the fields and the coupling constants, so that the fields have
the dimension GeV$^{1/2}$ and the couplings $g_3^2$, $\lambda_3$
have the dimension GeV.
Due to the convention $g'^2\sim g^3$,
we can use $h_3'=g_3'^2/4$, neglect the quartic
coupling of $B_0$-fields,
and use the indicated tree-level values $g_3$, $g_3'$
in the part mixing $A_0$ and $B_0$.
The problem is to calculate the parameters in eq.~\nr{action}
to order~$g^4$ in the coupling constants.

First, let us calculate how the 3d fields are related to the 4d fields.
The momentum-dependent contribution of the superheavy modes
to the two-point scalar correlator is
\be
{\cal Z}^{\phi}={\cal Z}^{\phi}_{\rm CT}-
\frac{3}{4}g^2{\cal Z}^{\phi}_{\rm SV}+
\frac{3}{2}g_Y^2{\cal Z}^{\phi}_{\rm FF}=
\frac{k^2}{16\pi^2}(-\frac{9}{4}g^2L_b+3g_Y^2L_f),
\ee
where the ${\cal Z}^{\phi}$'s are from eqs.~\nr{zfct}-\nr{zfff}.
For the spatial and temporal components of the gauge fields one gets
\be
{\cal Z}^{A}={\cal Z}^{A}_{\rm CT}-\frac{N_s}{2}g^2{\cal Z}^{A}_{\rm SS}+
2g^2{\cal Z}^{A}_{\eta\eta}
-2n_Fg^2{\cal Z}^{A}_{\rm FF}-g^2{\cal Z}^{A}_{\rm VV},
\ee
where the ${\cal Z}^{A}$'s are from eqs.~\nr{za0ct}-\nr{zaivv}.
This gives
\ba
{\cal Z}^{A_0} & = &
\frac{g^2k^2}{16\pi^2}\biggl[
\frac{4n_F}{3}(L_f-1)+\frac{N_s}{6}(L_b+2)-\frac{13}{3}L_b +\frac{8}{3}
\biggr],
\\ & & \nonumber \\
{\cal Z}^{A_i} & = &
\biggl(\delta_{ij}-\frac{k_ik_j}{k^2}\biggr)
\frac{g^2k^2}{16\pi^2}\biggl[
\frac{4n_F}{3}L_f+\frac{N_s}{6}L_b-\frac{13}{3}L_b -\frac{2}{3}
\biggr].
\ea
The $\cal Z$'s here correspond just to $\overline{\Pi}'(0)k^2$
in eq.~\nr{pik2}.
Hence the wave functions in the 3d action are related to
the renormalized 4d wave functions in the $\overline{\rm MS}$-scheme by
\ba
\phi_3^2 & = &
\frac{1}{T}
\phi^2 \biggl\{1+\frac{1}{16\pi^2}\biggl[-\frac{9}{4}g^2L_b+
3g_Y^2L_f\biggr]\biggr\}, \label{phi3} \\
\bigl(A_0^{3d}\bigr)^2 & = &
\frac{1}{T}
\bigl(A_0\bigr)^2\biggl\{1+\frac{g^2}{16\pi^2}\biggl[
\frac{4n_F}{3}(L_f-1)+\frac{N_s}{6}(L_b+2)-\frac{13}{3}L_b +\frac{8}{3}
\biggr]\biggr\}, \label{A03} \\
\bigl(A_i^{3d}\bigr)^2 & = &
\frac{1}{T}
\bigl(A_i\bigr)^2\biggl\{1+\frac{g^2}{16\pi^2}\biggl[
\frac{4n_F}{3}L_f+\frac{N_s}{6}L_b-\frac{13}{3}L_b -\frac{2}{3}
\biggr]\biggr\}. \label{Ai3}
\ea
The factor $1/T$ arises because of the rescaling in eq.~\nr{action}.
The loop corrections to the normalization of $B_0$ are of higher order
according to our convention. When the running of fields in 4d
is taken into account, $\phi_3$, $A_0^{3d}$
and $A_i^{3d}$ are seen to be independent of $\mu$.

The constant~$-2/3$ inside the square brackets
in the formulas for $A_0^{3d}$ and $A_i^{3d}$
in eqs.~\nr{A03}, \nr{Ai3}
is missing in~\cite{FKRS1}, due to
an error in eq.~(6.3) in~\cite{L}. This error propagates
to~$g_3^2$ and~$h_3$; the correct result for $g_3^2$ is
also given in eq.~(6) of~\cite{HL}.

Second, let us calculate the 1-loop corrections to
the coupling constants of the gauge sector.
The couplings $g_3^2$ and $h_3$
can be extracted from the $n\neq 0$ contributions to
the $(\phi\phi A_iA_j)$- and $(\phi\phi A_0A_0)$-correlators
at vanishing external momenta, respectively.
The corrections to the $(\phi\phi B B)$-
and $(\phi\phi B_0A_0)$-vertices are of higher order.
The contributions from the relevant diagrams
are in eqs.~\nr{ga00}-\nr{gaiffff}.
The coefficients are such that
\ba
{\cal G}^{A} & = &
{\cal G}^{A}_{0}
+{\cal G}^{A}_{\rm CT}
-6\lambda g^2{\cal G}^{A}_{\rm SS}
-g^4{\cal G}^{A}_{\rm SV}
-g^4{\cal G}^{A}_{\rm VV} \nonumber \\
& &
+6\lambda g^2{\cal G}^{A}_{\rm SSS}
+2g^4{\cal G}^{A}_{\rm VVV}
-3g^2g_Y^2{\cal G}^{A}_{\rm FFFF}.
\ea
This gives the effective vertices
\ba
& & \frac{g^2}{8}\phi_i\phi_iA^a_jA^a_j\biggl[
1+\frac{1}{16\pi^2}
\biggl(\frac{3}{4}g^2L_b+3g_Y^2L_f\biggr)\biggr],\nonumber \\
& & \frac{g^2}{8}\phi_i\phi_iA^a_0A^a_0\biggl[
1+\frac{1}{16\pi^2}
\biggl(\frac{3}{4}g^2L_b+3g_Y^2L_f
+\frac{23}{2}g^2+12\lambda-6g_Y^2
\biggr)
\biggr],
\ea
where the fields are those of the 4d theory.
When the fields are redefined according
to eqs.~\nr{phi3}-\nr{Ai3} and the vertex
is identified with the corresponding vertex in eq.~\nr{action},
one gets the final result for
the coupling constants $g_3^2$ and $h_3$:
\ba
g_3^2 & = & g^2(\mu)T\biggl[1+\frac{g^2}{16\pi^2}\biggl(
\frac{44-N_s}{6} L_b-
\frac{4n_F}{3} L_f+\frac{2}{3}\biggr)\biggr], \label{g32}\\
h_3 & = & \frac{1}{4}g^2(\mu)T
\biggl[1+ \frac{g^2}{16\pi^2}\biggl(
\frac{44-N_s}{6} L_b-
\frac{4n_F}{3} L_f
+\frac{53}{6}\nonumber \\
& & -\frac{N_s}{3}+\frac{4n_F}{3}
+\frac{3}{2}h^2-3t^2\biggr)\biggr].
\ea

As to the Higgs sector,
the 1-loop unresummed contribution to
the effective potential in Landau gauge is
\be
V_1(\varphi)={\cal C}_S(m_1)+3{\cal C}_S(m_2)+
2{\cal C}_V(m_T)+
{\cal C}_V(\sqrt{m_T^2+m_T'^2})+3{\cal C}_F(m_f),
\ee
where the ${\cal C}$'s are from
eqs.~\nr{cs}-\nr{cf}.
The masses appearing in $V_1(\varphi)$ are
\ba
&  & m_1^2 = -\nu^2+3\lambda\varphi^2,\quad
m_2^2=-\nu^2+\lambda\varphi^2,\nonumber \\
&  & m_T^2 = \frac{1}{4}g^2\varphi^2,\quad
m_T'^2=\frac{1}{4}g'^2\varphi^2,\quad
m_f^2=\frac{1}{2}g_Y^2\varphi^2. \label{4dms}
\ea
{}From the term quartic in masses in $V_1(\varphi)$,
one gets the $n\neq0$ contribution to the four-point scalar
correlator at vanishing momenta. Redefining the
field $\phi$ according to eq.~\nr{phi3},
the coupling constant $\lambda_3$ is then
\ba
\lambda_3 & = & \lambda(\mu)T\biggl\{
1-\frac{3}{4}\frac{g^2}{16\pi^2}\biggl[\biggl(
\frac{6}{h^2} -6+2h^2 \biggr)L_b+
\biggl(4 t^2-8\frac{t^4}{h^2}\biggr)L_f-
\frac{4}{h^2} \biggr]\biggr\}.
\ea
The coefficient of $\varphi^2/2$ in $V_1(\varphi)$ gives the 1-loop
result for the scalar mass squared. The result
is the term of order $g^2$ on the first line
of eq.~\nr{m32}.

For the 2-loop contribution to the mass squared $m_3^2$,
one needs the 2-loop effective potential $V_2(\varphi)$.
The diagrams needed for $V_2(\varphi)$ in the Standard Model
are those in Fig.~23 of~\cite{AE},
added by two purely scalar diagrams,
the figure-8 and the sunset. In terms of
eqs.~\nr{fae}-\nr{lae}, the result is
\ba
V_2(\varphi) & = &
-3\lambda^2\varphi^2\Bigl[{\cal D}_{\rm SSS}(m_1,m_1,m_1)
+{\cal D}_{\rm SSS}(m_1,m_2,m_2)\Bigr]
\nonumber \\ & &
-\frac{3}{8}g^2\Bigl[
{\cal D}_{\rm SSV}(m_1,m_2,m_T)
+{\cal D}_{\rm SSV}(m_2,m_2,m_T)\Bigr]
\nonumber \\ & &
-\frac{3}{64}g^4\varphi^2{\cal D}_{\rm SVV}(m_1,m_T,m_T)
-\frac{1}{2}g^2{\cal D}_{\rm VVV}(m_T,m_T,m_T)
\nonumber \\ & &
-\frac{3}{4}g_Y^2\Bigl[
{\cal D}_{\rm FFS}(m_f,m_f,m_1)
+{\cal D}_{\rm FFS}(m_f,m_f,m_2)
+2{\cal D}_{\rm FFS}(m_f,0,m_2)\Bigr]
\nonumber \\ & &
-\frac{3}{8}g^2\Bigl[
{\cal D}_{\rm FFV}(m_f,m_f,m_T)
+4{\cal D}_{\rm FFV}(m_f,0,m_T)
+(8n_F-5){\cal D}_{\rm FFV}(0,0,m_T)\Bigr]
\nonumber \\ & &
-3g^2{\cal D}_{\eta\eta {\rm V}}(m_T)
-4g_S^2{\cal D}_{\rm FFV}(m_f,m_f,0)
\nonumber \\ & &
-\frac{3}{4}\lambda\Bigl[
{\cal D}_{\rm SS}(m_1,m_1)
+2{\cal D}_{\rm SS}(m_1,m_2)
+5{\cal D}_{\rm SS}(m_2,m_2)\Bigr]
\nonumber \\ & &
-\frac{3}{16}g^2\Bigl[
{\cal D}_{\rm SV}(m_1,m_T)
+3{\cal D}_{\rm SV}(m_2,m_T)\Bigr]
-\frac{3}{4}g^2{\cal D}_{\rm VV}(m_T,m_T)
\nonumber \\ & &
+\frac{1}{2}{\cal D}_{\rm S}(m_1)
+\frac{3}{2}{\cal D}_{\rm S}(m_2)
+\frac{3}{2}{\cal D}_{\rm V}(m_T)
+3{\cal D}_{\rm F}(m_f),
\label{V2sum}
\ea
where the mass counterterms needed
for calculating ${\cal D}_{\rm S}$,
${\cal D}_{\rm V}$, and ${\cal D}_{\rm F}$
include those generated by the shift.
The $\imath_\epsilon$'s from the different
diagrams cancel in the sum. The linear terms
of the form $T^2 I_3(m)$ are cancelled by the thermal
counterterms as explained after eq.~\nr{thctc}.
Apart from vacuum terms,
the terms with $1/\epsilon$'s combine to
$\varphi^2/2$ multiplied by
\be
-\frac{T^2}{16\pi^2}\frac{\mu^{-4\epsilon}}{4\epsilon}
\biggl(\frac{81}{16}g^4+9\lambda g^2-12\lambda^2\biggr),\label{premct}
\ee
which is the order $g^4$-result for the mass counterterm of the 3d theory.
However, one can add higher-order corrections to
the mass counterterm by calculating the mass divergence
directly in the 3d theory, obtaining
\be
\delta m_3^2=-\frac{1}{16\pi^2}\frac{\mu^{-4\epsilon}}{4\epsilon}
\biggl(\frac{39}{16}g_3^4+12h_3g_3^2-6h_3^2
+9\lambda_3 g_3^2-12\lambda_3^2\biggr).\label{mct}
\ee
This agrees to order $g^4$ with eq.~\nr{premct}
and is the final result for $\delta m_3^2$.

Summing the finite contributions in eq.~\nr{V2sum},
one gets the expression for the renormalized part $m_3^2(\mu)$
of the mass squared. With the shorthand notations
\ba
\tilde{\nu}^2 & = & \nu^2(\mu)\biggl\{1- \frac{3}{4}\frac{g^2}{16\pi^2}
\biggl[\biggl(
h^2-3\biggr)L_b+2 t^2 L_f\biggr]\biggr\}, \\
\tilde{g}_Y^2 & = & T g_Y^2(\mu)\biggl\{1-\frac{3}{8}
\frac{g^2}{16\pi^2}\biggl[\biggl(6 t^2-
6-\frac{64}{3}s^2\biggr)L_f\nonumber \\
& + & 2+28 \ln{2}-12h^2\ln 2+
8 t^2\ln 2-\frac{64}{9}s^2(4\ln 2-3)\biggr]\biggr\} ,
\ea
the result after redefinition of fields is
\ba
m_3^2(\mu) & = & -\tilde{\nu}^2
+T\biggl(\frac{1}{2}\lambda_3+\frac{3}{16}g_3^2+\frac{1}{16}g_3'^2+
\frac{1}{4}\tilde{g}_Y^2\biggr)
\nonumber \\
& + &
\frac{T^2}{16\pi^2} \biggl[
g^4\biggl(\frac{137}{96}+\frac{3n_F}{2} \ln{2}+\frac{n_F}{12}\biggr)+
\frac{3}{4}\lambda g^2 \biggr]
\nonumber \\
& + &
\frac{1}{16\pi^2}
\biggl(\frac{39}{16}g_3^4+12h_3g_3^2-6h_3^2+
9\lambda_3g_3^2-12\lambda_3^2\biggr)
\biggl(\ln\frac{3 T}{\mu}+c\biggr).
\label{m32}
\ea
Here, as in eq.~\nr{mct}, we have taken into account higher-order
corrections in the logarithmic term.

The parameters $m_D'^2$, $m_D^2$ and $\lambda_A$ require
the calculation of the effect of $n\neq 0$ modes on the
$(B_0B_0)$-, $(A_0A_0)$- and $(A_0A_0A_0A_0)$-correlators
at vanishing momenta.
Using eqs.~\nr{a2s}-\nr{a2ff}, the
two-point correlator for the U(1)-field $B_0$ is
\ba
{\cal A}^{(2)}_{\rm U(1)} & = &
g'^2\Bigl[N_s{\cal A}^{(2)}_{\rm S}
-\frac{N_s}{2}{\cal A}^{(2)}_{\rm SS}
-\frac{10n_F}{3}{\cal A}^{(2)}_{\rm FF}\Bigr].\label{a2u1}
\ea
For the SU(2)-field $A_0$, one gets
\ba
{\cal A}^{(2)}_{\rm SU(2)} & = &
g^2\Bigl[N_s{\cal A}^{(2)}_{\rm S}
+{\cal A}^{(2)}_{\rm V}
-\frac{N_s}{2}{\cal A}^{(2)}_{\rm SS}
-{\cal A}^{(2)}_{\rm VV}
+2{\cal A}^{(2)}_{\eta\eta}
-2n_F{\cal A}^{(2)}_{\rm FF}\Bigr].
\ea
Using eqs.~\nr{a4ss}-\nr{a4ffff}, the
four-point correlator is
\ba
{\cal A}^{(4)}_{\rm SU(2)} & = &g^4\Bigl[
-\frac{N_s}{4}{\cal A}^{(4)}_{\rm SS}
+\frac{N_s}{2}{\cal A}^{(4)}_{\rm SSS}
-\frac{N_s}{8}{\cal A}^{(4)}_{\rm SSSS}
-\frac{1}{2}{\cal A}^{(4)}_{\rm VV} \nonumber \\
& &
+2{\cal A}^{(4)}_{\rm VVV}
-{\cal A}^{(4)}_{\rm VVVV}
+2{\cal A}^{(4)}_{\eta\eta\eta\eta}
+\frac{n_F}{2}{\cal A}^{(4)}_{\rm FFFF}\Bigr].\label{a4su2}
\ea
Since there is no tree-level term corresponding to
the correlators in eqs.~\nr{a2u1}-\nr{a4su2},
the redefinition of fields in eq.~\nr{A03} produces
terms of higher order. The final results
can then be read directly from eqs.~\nr{a2u1}-\nr{a4su2}:
\ba
m_D'^2 & = & \biggl(\frac{N_s}{6}+\frac{5n_F}{9}\biggr) g'^2 T^2, \\
m_D^2 & = & \biggl(\frac{2}{3}+\frac{N_s}{6}+\frac{n_F}{3}\biggr) g^2 T^2, \\
\lambda_A & = & T\frac{g^4}{16\pi^2} \frac{16+N_s-4 n_F}{3}.
\ea
In principle, the mass $m_D^2$ should be determined to
order $g^4$ to be compatible with the accuracy of
vacuum renormalization. We have, however, not made this calculation,
since the effect of $g^4$-corrections to $m_D^2$
contributes in higher order
than $g^4$ to the Higgs field effective potential $V(\varphi)$,
which drives the EW phase transition.

Using eqs.~\nr{run1}-\nr{run4}, one
sees that the quantities $g_3^2$, $h_3$,
$\lambda_3$, $\tilde{\nu}^2$ and $\tilde{g}_Y^2$
are independent of $\mu$ to the
order they are presented above. In other words,
when the running parameters
$g^2(\mu)$, $\nu^2(\mu)$, $\lambda(\mu)$
and $g_Y^2(\mu)$ are expressed in terms of physical
parameters in Sec.~\ref{MSbar-Ph}, the $\mu$-dependence
cancels in the 3d parameters.
The $\mu$-dependence of $\lambda_A$, $m_D^2$ is of higher order,
as well; actually $m_D^2$ runs only at order $g^6$~\cite{FKRS1}.
Note also that the $\mu$ in $m_3^2(\mu)$ is independent of the
$\mu$ used in the construction of the 3d theory (although the
notation is the same), since the bare mass parameter
$m_3^2(\mu)+\delta m_3^2$,
being the sum of eqs.~\nr{mct} and \nr{m32}, is independent of $\mu$.

\subsection{Integration over the heavy scale}
\label{iohs}

The action in eq.~\nr{action} can be further simplified
by integrating out the $A_0$- and $B_0$-fields.
The masses of these fields are of the order $gT$ and
$g'T\sim g^{3/2}T$, respectively.
The resulting action is of the form in eq.~\nr{action}
with the $A_0$- and $B_0$-fields left out and the parameters modified.
There are no infrared divergences related to this integration, either,
since the $A_0$- and $B_0$-fields are massive.
We denote the new parameters by a bar.
For the SU(2)-Higgs model, the relations of
the old and the barred parameters have been given
in~\cite{FKRS1}.

The calculation of the barred parameters proceeds
in complete analogy with dimensional reduction.
At 1-loop level, there is no momentum-dependent
correction from the heavy $A_0$- and $B_0$-fields
to the $\phi_3$-correlator,
so that $\bar{\phi}_3=\phi_3$. The momentum-dependent
correction to the $A_i$-correlator is
\be
-g_3^2{\cal Z}^{A_i}_{\rm LL},
\ee
where ${\cal Z}^{A_i}_{\rm LL}$ is from eq.~\nr{zaill}, so that
\be
\bigl(\bar{A}_i\bigr)^2 = \bigl(A_i^{3d}\bigr)^2
\biggl(1+\frac{g_3^2}{24\pi m_D}\biggr).
\ee
The field $B_0$ does not get normalized, since $B_0$ and $B_i$
do not interact.

To find $\bar{g}_3^2$, one can evaluate the two diagrams
with $A_0$ in the loop contributing to the
$(\phi_3)^2 \bigl(A^{3d}_{i}\bigr)^2$-vertex.
In terms of ${\cal G}_{\rm LL}$
and ${\cal G}_{\rm LLL}$ in eqs.~\nr{gll}-\nr{glll},
the result is
\be
g_3^2\delta_{ij}-8h_3g_3^2{\cal G}_{\rm LL}+
8h_3g_3^2{\cal G}_{\rm LLL}=
g_3^2\delta_{ij}.
\ee
Hence
\be
\bar{g}_3^2\bar{\phi}_3^2\bar{A}_i^2=
g_3^2\phi_3^2\bigl(A^{3d}_{i}\bigr)^2,
\ee
which gives
\be
\bar{g}_3^2=g_3^2\biggl(
1-\frac{g_3^2}{24\pi m_D}\biggr).
\ee
The coupling $g_3'^2$ does not get normalized,
since $B_0$ and $B_i$ do not interact.

The 1-loop corrections to the scalar coupling constant are
\be
-3h_3^2{\cal B}(0;m_D,m_D)
-\frac{1}{8}g_3^2g_3'^2{\cal B}(0;m_D,m_D')
-\frac{1}{16}g_3'^4{\cal B}(0;m_D',m_D').
\ee
This gives
\be
\bar{\lambda}_3=\lambda_3-\frac{1}{8\pi}\biggl(
3\frac{h_3^2}{m_D}+
\frac{g_3'^4}{16 m_D'}+\frac{g_3'^2 g_3^2}{4(m_D+m_D')}
\biggr).
\ee
The 1-loop corrections to the scalar mass parameter are
\be
3h_3\bar{I}_3(m_D)+\frac{1}{4}g_3'^2\bar{I}_3(m_D'),
\ee
giving the first line in eq.~\nr{bm32}.
To calculate the 2-loop corrections, one needs the
effective potential. The 2-loop contribution from
the heavy scale to the effective potential is
\ba
V_2^{\rm heavy}(\varphi) & = &
\frac{3}{2}h_3\Bigl[
D_{\rm LS}(m_1)+3D_{\rm LS}(m_2)\Bigr]
+3g_3^2D_{\rm LV}(m_T) \nonumber \\
& & -3h_3^2\varphi^2D_{\rm LLS}(m_1)
-\frac{3}{2}g_3^2 D_{\rm LLV}(m_T),
\ea
where the $D$'s are from eqs.~\nr{dls}-\nr{dllv}
and we used $g'^2\sim g^3$.
The $1/\epsilon$-parts modify the mass counterterm
of eq.~\nr{mct} to become
\be
-\frac{1}{16\pi^2}\frac{\mu^{-4\epsilon}}{4\epsilon}
\biggl(\frac{51}{16}g_3^4
+9\lambda_3 g_3^2-12\lambda_3^2\biggr).
\ee
However, one can again include higher order corrections
by calculating the counterterm directly in the final theory,
getting
\be
\delta\bar{m}_3^2=-\frac{1}{16\pi^2}\frac{\mu^{-4\epsilon}}{4\epsilon}
\biggl(\frac{51}{16}\bar{g}_3^4
+9\bar{\lambda}_3 \bar{g}_3^2-12\bar{\lambda}_3^2\biggr).\label{bcmt}
\ee
Finally, the renormalized
mass parameter $\bar{m}_3^2(\mu)$ is
\ba
\bar{m}_3^2(\mu) & = &m_3^2(\mu)-\frac{1}{4\pi}\biggl(3 h_3 m_D +
\frac{1}{4}g_3'^2m_D'\biggr) \nonumber \\
& &\hspace*{0.5cm} +\frac{1}{16\pi^2}\biggl[
\biggl(-\frac{3}{4}g_3^4+12h_3g_3^2-6h_3^2\biggr)
\ln\frac{\mu}{2m_D}+3h_3g_3^2-3h_3^2\biggr] \nonumber \\
& & \nonumber \\
& = & -\tilde{\nu}^2
+T\biggl(\frac{1}{2}\lambda_3+\frac{3}{16}g_3^2+\frac{1}{16}g_3'^2+
\frac{1}{4}\tilde{g}_Y^2\biggr)
-\frac{1}{4\pi}\biggl(3 h_3 m_D +
\frac{1}{4}g_3'^2m_D'\biggr) \nonumber \\
& & \hspace*{0.5cm}
+\frac{T^2}{16\pi^2} \biggl[
g^4\biggl(\frac{137}{96}+\frac{3n_F}{2} \ln{2}+\frac{n_F}{12}\biggr)+
\frac{3}{4}\lambda g^2\biggr]
\nonumber \\
& & \hspace*{0.5cm}
+\frac{1}{16\pi^2}
\biggl[
\biggl(-\frac{3}{4}g_3^4+12h_3g_3^2-6h_3^2
\biggr)\biggl(\ln\frac{3 T}{2 m_D}+c\biggr)
+3h_3g_3^2-3h_3^2
\biggr]
\nonumber \\
& & \hspace*{0.5cm}
+\frac{1}{16\pi^2}
\biggl(\frac{51}{16}\bar{g}_3^4+
9\bar{\lambda}_3\bar{g}_3^2-12\bar{\lambda}_3^2\biggr)
\biggl(\ln\frac{3 T}{\mu}+c\biggr),
\label{bm32}
\ea
where we used eq.~\nr{m32}
and included higher order corrections in the logarithmic
term on the last line.
Eq.~\nr{bm32} completes the evaluation of the couplings of
the 3d SU(2)$\times$U(1)+Higgs theory.

\section{Relation of the $\overline{\rm MS}$-scheme and Physics in
the Standard Model}
\label{MSbar-Ph}

In Sec.~\ref{DRinSM}, we have given the relations of the running parameters
in the $\msbar$-scheme to the parameters of the
effective 3d theory with accuracy $g^4$. For this accuracy to be
meaningful, the running parameters in the $\overline{\rm MS}$-scheme
should be expressed with accuracy $g^4$ in terms of true physical
parameters, like pole masses. We give these relations in the
present Section.

The accuracy $g^4$ requires 1-loop renormalization
of the vacuum theory. The 1-loop renormalization of the
Standard Model is, of course, a well studied subject.
Usually, however, one does not use the $\msbar$-scheme
we have employed above,
but the so called on-shell scheme~\cite{S1,S2,H1,H2}.
In the on-shell scheme, the divergences appearing in the
loop integrals are handled with dimensional regularization
as in the $\overline{\rm MS}$-scheme, but the finite parts
of the counterterms are chosen differently. Indeed,
the renormalized parameters are chosen to be physical
and independent of $\mu$. For instance, the renormalized
mass squared of the scalar field is $\nu^2_{\rm os}=m_H^2/2$,
and the gauge and scalar couplings are given by
\be
g^2_{\rm os}=\frac{e^2_{\rm os}}{s^2_W}, \quad
g'^2_{\rm os}=\frac{e^2_{\rm os}}{c^2_W}, \quad
\lambda_{\rm os}=\frac{g^2_{\rm os}}{8}\frac{m_H^2}{m_W^2}.\label{os}
\ee
Here $m_H$ and $m_W$ are the physical pole masses of the
Higgs particle and the W boson, and
\be
e^2_{\rm os}\equiv 4 \pi\alpha,\quad
c_W\equiv\frac{m_W}{m_Z},\quad
s_W\equiv\sqrt{1-\frac{m_W^2}{m_Z^2}},\label{notation}
\ee
where $\alpha$ is the electromagnetic fine structure constant
defined in the Thomson limit, and $m_Z$ is the physical pole mass
of the Z boson. The parameters $c_W$ and $s_W$ are just
shorthand notations without any higher order corrections.

To convert results from the on-shell scheme
to the $\overline{\rm MS}$-scheme, let us note that since
all physical quantities derived in the two schemes
are exactly the same, the bare Lagrangians, including
the counterterms, must be the same.
In particular, all the bare parameters of the theories
must be the same (often the wave function renormalization
factors are not even needed, see e.g.~\cite{S1}).
For the gauge coupling this means
\be
g_B^2=g^2(\mu)+\delta g^2(\mu)=g^2_{\rm os}+\delta g^2_{\rm os},
\ee
that is,
\be
g^2(\mu)=g^2_{\rm os}\biggl(1+\frac{\delta g^2_{\rm os}-
\delta g^2(\mu)}{g^2_{\rm os}}\biggr)
\label{g2mu}.
\ee
The counterterm $\delta g^2(\mu)$ of the $\overline{\rm MS}$-scheme
cancels the $1/\epsilon$-parts in $\delta g^2_{\rm os}$, so that
eq.~\nr{g2mu} is finite. In this way, the running parameters
in the $\overline{\rm MS}$-scheme can be expressed
in terms of physical parameters. We shall work out the
explicit expressions in some detail below; the reader
only interested in the final numerical results
for the 3d parameters in terms of the
physical 4d parameters may turn to Sec.~\ref{results}.

\subsection{The parameters $g^2(\mu)$ and $g'^2$ in terms of
physical parameters}
\label{ggprim}

The expression for $\delta g^2_{\rm os}$, determining
$g^2(\mu)$ through eq.~\nr{g2mu}, can in principle
be read directly for example from eq.~(28.a) in~\cite{S1}.
With our sign conventions, the equation reads
\be
\frac{\delta g^2_{\rm os}}{g^2_{\rm os}}=
2\frac{\delta e_{\rm os}}{e_{\rm os}}-
\frac{c_W^2}{s_W^2}\biggl[\frac{\tilde{\Pi}_Z(-m_Z^2)}{m_Z^2}-
\frac{\tilde{\Pi}_W(-m_W^2)}{m_W^2}\biggr]. \label{dg2}
\ee
Here $\tilde{\Pi}$ means the unrenormalized but regularized self-energy.
This equation is gauge-independent, since $\delta e_{\rm os}$
and the self-energies at the pole are~\cite{S1}.
However, a reliable estimate of eq.~\nr{dg2}
cannot be given purely perturbatively, since
the expression for $\delta e_{\rm os}$ contains
the photon self-energy $\tilde{\Pi}_{\gamma}(k^2)/k^2$ evaluated
at vanishing momentum $k^2$. Indeed, $\tilde{\Pi}_{\gamma}(k^2)/k^2$
includes logarithms of all the small lepton and quark masses,
as is seen e.g. in eq.~(5.40) of~\cite{H1}. This
indicates that strong interactions are important for $\delta e_{\rm os}$.
There exists a standard technique of expressing
the hadronic contribution to $\delta e_{\rm os}$
in terms of a dispersion relation, and hence
$\delta e_{\rm os}$ is known with very good accuracy~\cite{H2,C}
in spite of strong interactions. We find it convenient, though,
to write the expression for $\delta g^2_{\rm os}$ in
a form somewhat different from eq.~\nr{dg2}, so that $\delta e_{\rm os}$ is
not directly visible. Such a way is expressing
$\delta g^2_{\rm os}$ in terms of the Fermi constant $G_\mu$.

The Fermi constant $G_\mu$ is defined~\cite{S1,S2,C}
in terms of the muon lifetime by
\be
\frac{1}{\tau_\mu}=\frac{G_\mu^2m_\mu^5}{192 \pi^3}
\biggl(1-8\frac{m_e^2}{m_\mu^2}\biggr)
\biggl[1+\frac{\alpha}{2\pi}
\biggl(1+\frac{2\alpha}{3\pi}\ln\frac{m_\mu}{m_e}\biggl)
\biggl(\frac{25}{4}-\pi^2\biggr)\biggr].\label{Gmu}
\ee
The $\alpha$-corrections here account for the QED-corrections to
muon decay in the local Fermi-model. On the other hand,
calculating the muon lifetime in the Standard Model, one
gets a prediction for the $G_\mu$ of eq.~\nr{Gmu}~\cite{S1,S2,H1,H2}:
\be
\frac{G_\mu}{\sqrt{2}}=\frac{g^2_{\rm os}}{8 m_W^2}\frac{1}{1-\Delta r}.
\label{Gu}
\ee
Here $\Delta r$ depends on the parameters of the Standard Model.
Using eqs.~(28.a),~(34.b) of~\cite{S1},
the 1-loop expression for $\Delta r$ in the on-shell scheme
in the Feynman-$R_\xi$-gauge can be written as
\ba
\Delta r & = & \frac{{\mathop {\rm Re}}
\Bigl[\tilde{\Pi}_W(0)-\tilde{\Pi}_W(-m_W^2)\Bigr]}{m_W^2}+
\frac{\delta g^2_{\rm os}}{g^2_{\rm os}} \nonumber \\
& + & \frac{g^2_{\rm os}}{16 \pi^2}
\biggl[
4\biggl(\frac{1}{\epsilon}+\ln\frac{\mu^2}{m_W^2}\biggr)
+6+\frac{7-4 s_W^2}{2s_W^2}\ln c_W^2\biggr].
\label{Dr}
\ea
This equation can also be extracted from~\cite{H2} as a combination of
eqs.~(3.8), (3.16), (3.17), (4.18), (A.1), (A.2) and (B.3).
Let us note that the value of eq.~\nr{Dr}
is finite and gauge-independent, since~$\Delta r$
is a physical observable.
In addition, $\delta g^2_{\rm os}/g^2_{\rm os}$ and
$\tilde{\Pi}_W(-m_W^2)$ are gauge-independent. However,
$\tilde{\Pi}_W(0)$ is not gauge-independent, and eq.~\nr{Dr}
as a whole holds only in the Feynman-$R_\xi$ gauge.

Usually eq.~\nr{Dr}, with $\delta g^2_{\rm os}$
plugged in from eq.~\nr{dg2} and $g^2_{\rm os}$ from
eqs.~\nr{os},~\nr{notation},
is used in determining $m_W$ from eq.~\nr{Gu}
in terms of the very precisely known
parameters $G_\mu$ and $m_Z$,
and the masses $m_t$ and $m_H$~\cite{H2,C}.
For fixed $m_t$ and $m_H$, the estimated uncertainty
in the value of $m_W$ determined this way is only of
the order of $0.01$~GeV~\cite{H2,C,SIII,K}.
This is much smaller than the present experimental uncertainty
in the W mass, $m_W=80.22\pm 0.18$~GeV~\cite{C}.
In other words, with the present accuracy in the determination
of $m_W$, one should replace $m_W$ as an input parameter
with the hadronic contribution to $\delta e_{\rm os}$.
The $m_W$ obtained this way is shown in Table~\ref{WHrel}
as a function of $m_t$ and $m_H$ for $\alpha_S=0.125$.

\begin{table}[htbp]
\centering

\begin{minipage}[t]{13cm}
\caption[a]{\protect
The mass $m_W$ as a function of $m_t$ and $m_H$ according to~\cite{T}.
The uncertainty in $m_W$ is about $0.01$~GeV for fixed
$m_t$, $m_H$~\cite{H2,C,SIII,K}.}
\vspace*{4mm}
\end{minipage}

\begin{tabular}{|c|c|c|c|c|c|c|c|c|c|} \hline
$m_t\backslash m_H$ &
35    & 50    & 60    & 70    & 80    & 90    & 100   & 200   & 300  \\ \hline
165 &
80.38 & 80.36 & 80.35 & 80.34 & 80.34 & 80.33 & 80.32 & 80.28 & 80.25 \\ \hline
175 &
80.44 & 80.42 & 80.41 & 80.41 & 80.40 & 80.39 & 80.39 & 80.34 & 80.31 \\ \hline
185 &
80.50 & 80.49 & 80.48 & 80.47 & 80.46 & 80.46 & 80.45 & 80.40 & 80.37 \\ \hline
\end{tabular}
\label{WHrel}
\end{table}

When $m_W$ is fixed, either from Table~1
or in the future from experiment,
$\Delta r$ is known from eq.~\nr{Gu},
and $\delta g^2_{\rm os}$
can be solved from eq.~\nr{Dr} in a simple form.
Using eq.~\nr{g2mu}, one then gets
\ba
g^2(\mu) & = & g^2_0\biggl\{1+
\frac{g^2_0}{16 \pi^2}
\biggl[\biggl(\frac{4 n_F}{3}-\frac{43}{6}\biggr)
\ln\frac{\mu^2}{m_W^2} \nonumber\\
& - & \frac{33}{4} F(m_W;m_W,m_W)+
\frac{1}{12}(h^4-4h^2+12)F(m_W;m_W,m_H) \nonumber\\
& - &
\frac{1}{2}(t^4+t^2-2)F(m_W;m_t,0)
-2\ln{t}-\frac{h^2}{24} +\frac{t^2}{4} +\frac{20 n_F}{9}-
\frac{257}{72}\biggr] \biggr\}.\label{fineq}
\ea
Here we used $g'^2\sim g^3$, and
defined $g_0^2=4\sqrt{2}G_\mu m_W^2=g^2_{\rm os}/(1-\Delta r)$.
The reason for using $g_0^2$ as the tree-level value instead of
$g^2_{\rm os}$ is that for $g^2_{\rm os}$ there would be a
rather large correction $\Delta r$
in the 1-loop formula, indicating bad convergence.
For instance, for $m_W=80.22$~GeV, $\Delta r =0.045$.
The physical reason~\cite{S2,H2,C} for
the large correction is that $g^2_{\rm os}$ is defined
in terms of the fine structure constant $\alpha$ measured at vanishing
momentum scale, whereas the momentum scale of weak interactions
is~$m_W^2$. With~$g_0^2$,
the 1-loop correction is extremely small for $\mu\sim m_W$.

The function $F(k;m_1,m_2)$ in eq.~\nr{fineq}
has been defined in~\cite{H1},
and its explicit form is given in eq.~\nr{Fkm}.
Note that from eqs.~\nr{Dr} and~\nr{fineq} one can see
that there are no dangerous logarithms in $\tilde{\Pi}_W(0)/m_W^2$
in contrast to $\tilde{\Pi}_\gamma(0)/k^2$, since
any such logarithms are suppressed by $m_f^2/m_W^2$.
Eq.~\nr{fineq} is the final result for $g^2(\mu)$
in terms of physical parameters.

In analogy with the definition for $g_0^2$, we define the U(1)
coupling to be
\be
g'^2=g_0^2\frac{s_W^2}{c_W^2}=g_0^2\frac{m_Z^2-m_W^2}{m_W^2} \label{gprim}
\ee
instead of $g'^2_{\rm os}$. The value of $g'^2$ is
larger than the value of $g'^2_{\rm os}$, corresponding
again roughly to $\alpha_{\rm EM}$ running from the Thomson limit $q^2=0$
to the electroweak scale.

\subsection{The parameters $\nu^2(\mu)$, $\lambda(\mu)$,
and $g_Y^2(\mu)$ in terms of physical parameters}
\label{nulgY}

The parameters $\nu^2(\mu)$, $\lambda(\mu)$,
and $g_Y^2(\mu)$ could be determined from the precision
calculations in the on-shell scheme just as $g^2(\mu)$.
For illustration, however, we will calculate the 1-loop corrections
to the tree-level values
\be
\nu^2=\frac{1}{2}m_H^2,\quad
\lambda=\frac{1}{\sqrt{2}}G_\mu m_H^2=\frac{g_0^2}{8}
\frac{m_H^2}{m_W^2},\quad
g_Y^2=2\sqrt{2}G_\mu m_t^2=\frac{g_0^2}{2}\frac{m_t^2}{m_W^2} \label{0l}
\ee
in some more detail.

\begin{figure}[tbh]
\vspace*{-1cm}

\hspace*{-0.5cm}
\epsfysize=20cm
\epsffile[50 80 500 680]{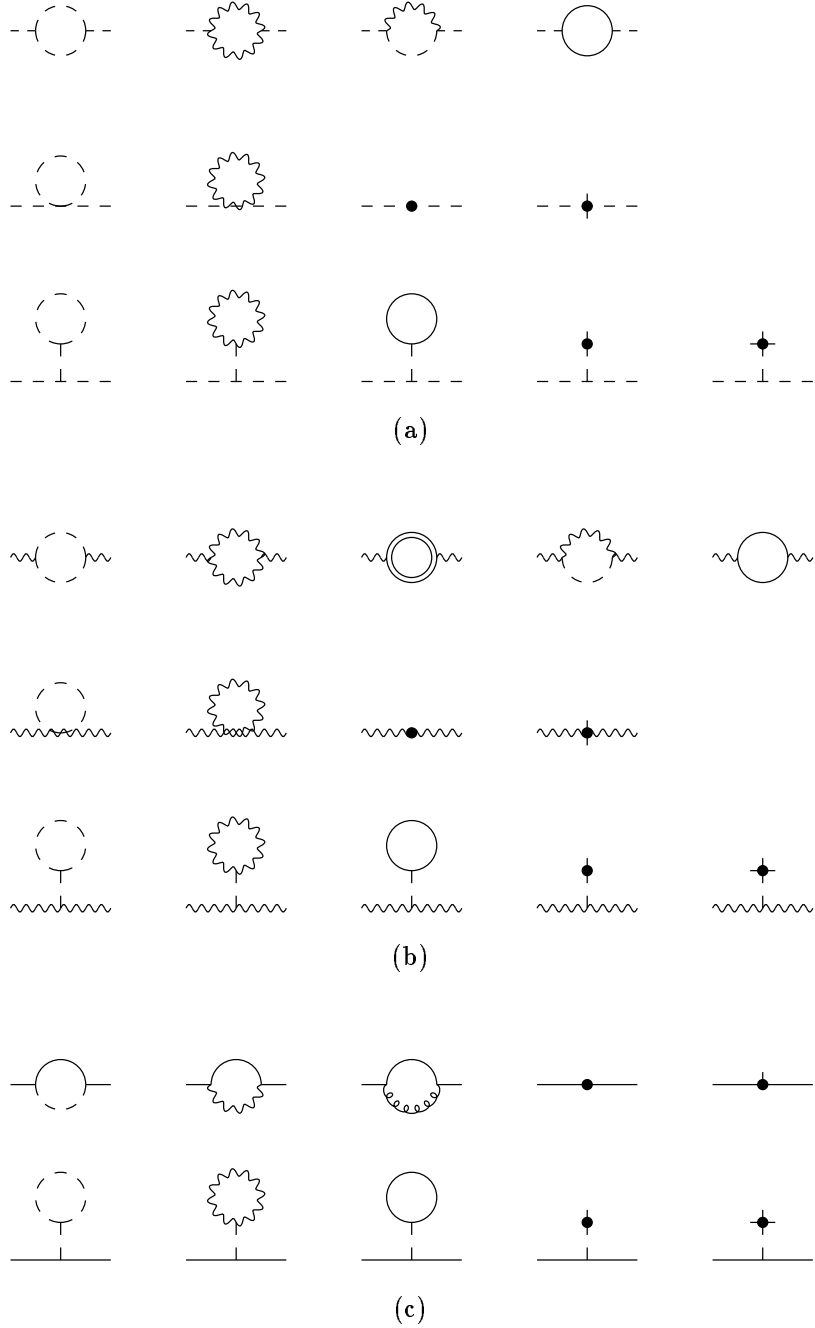}
\vspace*{-0.5cm}

\caption[a]{\protect
The diagrams needed for calculating the pole mass
of (a) the Higgs particle, (b) the W~boson and
(c) the top quark. Curly line is the gluon propagator.
For scalars, the counterterm indicated by the bare blob
contains both the mass counterterm $\delta\nu^2$
and the counterterm from wave function renormalization.
For vectors and fermions, the bare blob contains only the
counterterm from wave function renormalization.
The blobs with the short lines attached denote
counterterms generated by the shift.}
\label{pi}
\end{figure}

To calculate $\nu^2(\mu)$, $\lambda(\mu)$, and $g_Y^2(\mu)$
at 1-loop level, one has to calculate the 1-loop corrections
to the propagators of the Higgs particle, $W$-boson and
top quark in the broken phase, and extract from these
the pole masses. The diagrams needed are shown in Fig.~\ref{pi}.
To go to the broken phase, the Higgs field $\phi_1$
is shifted to the classical minimum $\varphi$,
where $\varphi^2=\nu^2/\lambda$.
The masses appearing in the Feynman rules are denoted
by $m_1^2=2\nu^2$ for the Higgs field,
$m_T^2=g^2\nu^2/4\lambda$ for the $W$-boson,
and $m_f^2=g_Y^2\nu^2/2\lambda$ for the top quark.
The radiatively corrected 1-loop propagators are then of the form
\ba
\langle\phi_1(-p)\phi_1(p)\rangle & = & \frac{1}{p^2+m_1^2-\Pi_H(p^2)},
\nonumber \\
\langle A^a_\mu(-p)A^b_\nu(p)\rangle & = &
\delta^{ab}\frac{\delta_{\mu\nu}-p_\mu p_\nu/p^2}
{p^2+m_T^2-\Pi_W(p^2)}
+ {\rm \quad longitudinal\; part} \label{prop} \\
\langle\Psi_\alpha(p)\overline{\Psi}_\beta(p)\rangle & = &
\biggl[\frac{1}{i\slash{p}+m_f+i\slash{p}\Sigma_v(p^2)+
i\slash{p}\gamma_5\Sigma_a(p^2)+
m_f\Sigma_s(p^2)}\biggr]_{\alpha\beta}. \nonumber
\ea

To give the results for the radiatively corrected propagators,
we use the function $F(k;m_1,m_2)$ defined in~\cite{H1}. For
$|m_1-m_2|<k<m_1+m_2$, $F(k;m_1,m_2)$ is
\ba
F(k;m_1,m_2) & = & 1-\frac{m_1^2-m_2^2}{k^2}\ln\frac{m_1}{m_2}+
\frac{m_1^2+m_2^2}{m_1^2-m_2^2}\ln\frac{m_1}{m_2} \label{Fkm} \\
& - & \frac{2}{k^2}\sqrt{(m_1+m_2)^2-
k^2}\sqrt{k^2-(m_1-m_2)^2}\arctan\frac{\sqrt{k^2-(m_1-m_2)^2}}
{\sqrt{(m_1+m_2)^2-k^2}},\nonumber
\ea
and has the special values
\ba
F(m_1;m_1,m_2) & = & 1-r^2\frac{3-r^2}{1-r^2}\ln{r}-
2 r \sqrt{4-r^2}\arctan\frac{\sqrt{2-r}}{\sqrt{2+r}}, \nonumber \\
F(m_1;m_2,m_2) & = & 2-2 \sqrt{4 r^2-1}\arctan\frac{1}{\sqrt{4 r^2-
1}}, \nonumber \\
F(m;m,m) & = & 2-\frac{\pi}{\sqrt{3}},
\ea
where $r=m_2/m_1$. We also need the analytical continuation
\be
F(m_1;m_2,0) = 1+(r^2-1)\ln\biggl(1-\frac{1}{r^2}\biggr).
\ee
For $m_H<2m_W$, the only formula needed in the region
where it develops an imaginary part, is $F(m_1;m_2,0)$.

With the help of $F(k;m_1,m_2)$, one can write down the special values
\ba
\Pi_H(-m_H^2) & = & \!\!
\frac{3}{8}\frac{g^2}{16 \pi^2}m_H^2
\biggl[2(h^2+2t^2-3)\ln\frac{\mu^2}{m_W^2} +
3 h^2 F(m_H;m_H,m_H) \nonumber \\
& + & \!\!\frac{h^4-4h^2+12}{h^2}F(m_H;m_W,m_W)-
4 t^2 \frac{4t^2-h^2}{h^2}F(m_H;m_t,m_t)\nonumber \\
& - & \!\!2 h^2 \ln{h}-8 t^2 \ln{t}-
2 h^2-2-12\frac{1}{h^2}+16 \frac{t^4}{h^2}\biggr],\label{PH}\\
\Pi_W(-m_W^2) & = & \frac{3}{8}\frac{g^2}{16 \pi^2}m_W^2
\biggl[ 2\Bigl( \frac{16 n_F-59}{9}-
h^2-2 t^2-6\frac{1}{h^2}+
8 \frac{t^4}{h^2}\Bigl)\ln\frac{\mu^2}{m_W^2} \nonumber\\
& - & 22 F(m_W;m_W,m_W)+
\frac{2}{9}(h^4-4h^2+12)F(m_W;m_W,m_H) \nonumber\\
& - &
\frac{4}{3}(t^4+t^2-2)F(m_W;m_t,0) \nonumber\\
& + & 4 h^2 \frac{h^2-2}{h^2-1} \ln{h}+
\frac{8}{3}\Bigl(3t^2-2-12\frac{t^4}{h^2}\Bigr)\ln{t}-
\frac{22}{9} h^2-\frac{4}{h^2}-\frac{4}{3} t^2 \nonumber\\
& + & 16 \frac{t^4}{h^2}+
\frac{4}{27}(40 n_F-17)+
\frac{8}{3}\Bigl(1-\frac{4}{3} n_F\Bigr)\ln(-1-i\epsilon)\biggr], \label{PW}\\
\Sigma_v(-m_t^2)-\Sigma_s(-m_t^2) & = &
\frac{3}{16}\frac{g^2}{16 \pi^2}
\biggl[2\Bigl(t^2-h^2-6\frac{1}{h^2}+
8\frac{t^4}{h^2}-\frac{32}{3}s^2
\Bigr)\ln\frac{\mu^2}{m_W^2}\nonumber\\
& + &\frac{2}{3}(4t^2-h^2)F(m_t;m_t,m_H)
+\frac{2}{3}\Bigl(1-\frac{1}{t^2}\Bigr)F(m_t;m_t,m_W)\nonumber \\
& + & \frac{2}{3}\Bigl(t^2+1-\frac{2}{t^2}\Bigr)F(m_t;m_W,0) \nonumber\\
& + & 4 h^2 \ln{h}-32 \frac{t^4}{h^2} \ln{t}-
\frac{4}{3}t^2\frac{2t^2+h^2}{t^2-h^2}\ln\frac{t}{h}+
\frac{128}{3}s^2\ln t \nonumber\\
& - & 2+2t^2-2h^2-\frac{4}{h^2}+16\frac{t^4}{h^2}
-\frac{256}{9}s^2\biggr]. \label{PF}
\ea
Here we again used notation from eq.~\nr{hts}.
The expressions~\nr{PH}-\nr{PF} are gauge-independent,
and are the only values of $\Pi_H$, $\Pi_W$ and the $\Sigma$'s
that will be needed here. The value of $\Pi_W(-m_W^2)$ can
be extracted from~\cite{H1}, and that
of $\Pi_H(-m_H^2)$ from~\cite{CEQR}. As a check, we have explicitly
verified the gauge-independence of $\Pi_W(-m_W^2)$.

Numerically the dominant terms in eqs.~\nr{PH}-\nr{PF}
are the fermionic contributions $16t^4/h^2$ and
$-32(t^4/h^2)\ln(\mu/m_t)$. These terms come from
the fermionic tadpoles, and from the fermionic loops in the
Higgs and $W$-boson correlators. The fermionic terms
are large because $m_t/m_W$ is large,
and hence the gauge coupling is large: $g_Y\approx 1.0$.
Consequently, higher order fermionic corrections
are rather important.

With eqs.~\nr{PH}-\nr{PF} and~\nr{fineq}, one can
express $\nu^2(\mu)$, $\lambda (\mu)$ and $g_Y^2(\mu)$
in terms of physical parameters. To do so, the physical
pole masses have to be extracted from eqs.~\nr{prop}.
For the Higgs particle and $W$-boson, this is
straightforward. For the top quark, the 1-loop
equation for the physical mass $m_t$ is
\be
\overline{u}(p)
\bigl[i\slash{p}+m_f+i\slash{p}\Sigma_v(-m_t^2)+
i\slash{p}\gamma_5\Sigma_a(-m_t^2)+
m_f\Sigma_s(-m_t^2)\bigr]u(p)=0. \label{teq}
\ee
Here $u(p)$ is an asymptotic spinor
satisfying $(i\slash{p}+m_t)u(p)=0$.
In eq.~\nr{teq}, the factor $\Sigma_a$ multiplying $\gamma_5$
does not affect the physical mass, since
$\bar{u}(p)\gamma_5 u(p)=0$. The factor $\Sigma_a$ would have an effect
if the top mass were determined from the requirement
that the determinant of the inverse propagator vanishes, in
which case $\Sigma_a$ produces an
unphysical imaginary part to the self-energy even for $m_t<m_W$.
Physically, the reason why the top quark can be considered
an asymptotic state, is that the time scale~$\tau\sim m_W^4/m_t^5$
of weak interactions is much smaller
than the scale $1$~fm of strong interactions.

Evaluating the expressions for the
Higgs particle, W boson, and top quark masses,
and expressing $m_1$, $m_T$ and $m_f$ in terms of the coupling constants,
one can then solve for the parameters $\nu^2(\mu)$,
$\lambda(\mu)$ and $g_Y^2(\mu)$. The results are
\ba
\nu^2(\mu) & = &\frac{m_H^2}{2}\,{\mathop{\rm Re}}
\biggl[1+\frac{\Pi_H(-m_H^2)}{m_H^2}\biggr], \nonumber \\
\lambda (\mu) & = & \frac{g_0^2}{8}\frac{m_H^2}{m_W^2}
\,{\mathop{\rm Re}} \biggl[1-
\frac{\Pi_W(-m_W^2)}{m_W^2}+\frac{\delta g^2(\mu)}{g_0^2}
+\frac{\Pi_H(-m_H^2)}{m_H^2}\biggr], \label{physMS} \\
g_Y^2(\mu) & = & \frac{g_0^2}{2}\frac{m_t^2}{m_W^2}
\,{\mathop{\rm Re}} \biggl[
1-\frac{\Pi_W(-m_W^2)}{m_W^2}+\frac{\delta g^2(\mu)}{g_0^2}
+2\Sigma_v(-m_t^2)-2\Sigma_s(-m_t^2)\biggr].
\nonumber
\ea
Here $\delta g^2(\mu)$ is the renormalized
1-loop correction in eq.~\nr{fineq}.
The $\mu$-dependences in eqs.~\nr{physMS}
naturally reproduce those in eqs.~\nr{run1}-\nr{run4}.
We will not write down explicitly the expressions
in eq.~\nr{physMS}, since
we did not find any significant simplification
in the final result, apart from the $\mu$-dependent terms.
Together with eqs.~\nr{fineq} and~\nr{gprim} and
the value $\alpha_S=0.125$, eqs.~\nr{physMS}
complete the relation of $\overline{\rm MS}$
to Physics.

\subsection{Numerical results for the parameters
of the effective 3d theory in Standard Model}
\label{results}

In Secs.~\ref{ggprim} and~\ref{nulgY} we have expressed
the five parameters $g^2(\mu)$, $g'^2$,
$\nu^2(\mu)$, $\lambda(\mu)$, and $g_Y^2(\mu)$ in terms
of the five physical parameters $G_\mu$, $m_W$, $m_Z$,
$m_H$ and $m_t$. In addition, due to the
large experimental uncertainty in $m_W$,
we replaced $m_W$ as an input
parameter with the hadronic contribution to the photon self-energy
through Table~1.
We have then four parameters left: the very well known
$G_\mu=1.16639\times 10^{-5}$ GeV$^{-2}$~\cite{PDG},
$m_Z=91.1887$~GeV~\cite{C}, the less well
known $m_t=175$ GeV~\cite{CDF}, and the
unknown $m_H$. We shall fix $m_t$ and use the
Higgs mass as a free parameter. Then
we can calculate the parameters of the effective
3d theory in terms of $m_H$ and $T$.

We do not write down the formulas
for the 3d parameters in terms of the physical 4d
parameters from eqs.~\nr{fineq},~\nr{PH}-\nr{PF},~\nr{physMS}
and Secs.~\ref{ioss},~\ref{iohs} explicitly,
since we found no significant simplification in the final result
apart from the $\ln\mu$-terms.

\begin{figure}[tbh]
\vspace*{-1cm}

\hspace*{-1.5cm}
\epsfysize=14cm
\epsffile{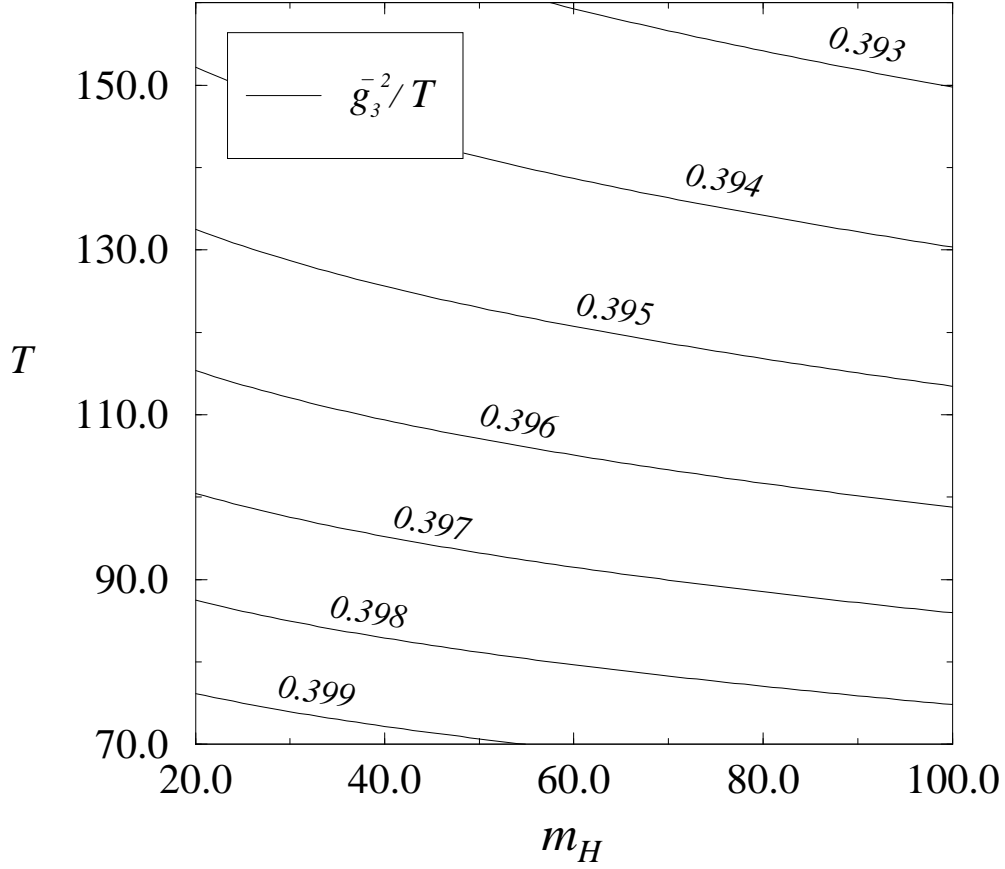}

\caption[a]{\protect
The scale $\bar{g}_3^2$ of the 3d
SU(2)+Higgs theory divided by the temperature $T$,
as a function of $m_H$ and $T$.
The dependence of $\bar{g}_3^2/T$
on $m_H$ is caused by 1-loop corrections
including the Higgs particle, and by the implicit
dependence of $m_W$ on $m_H$ through Table~1.
The dependence on temperature is caused by the logarithmic
running in eq.~\nr{g32}.}
\label{g3mHT}
\end{figure}

\begin{figure}[tbh]
\vspace*{-1cm}

\hspace*{-1.5cm}
\epsfysize=14cm
\epsffile{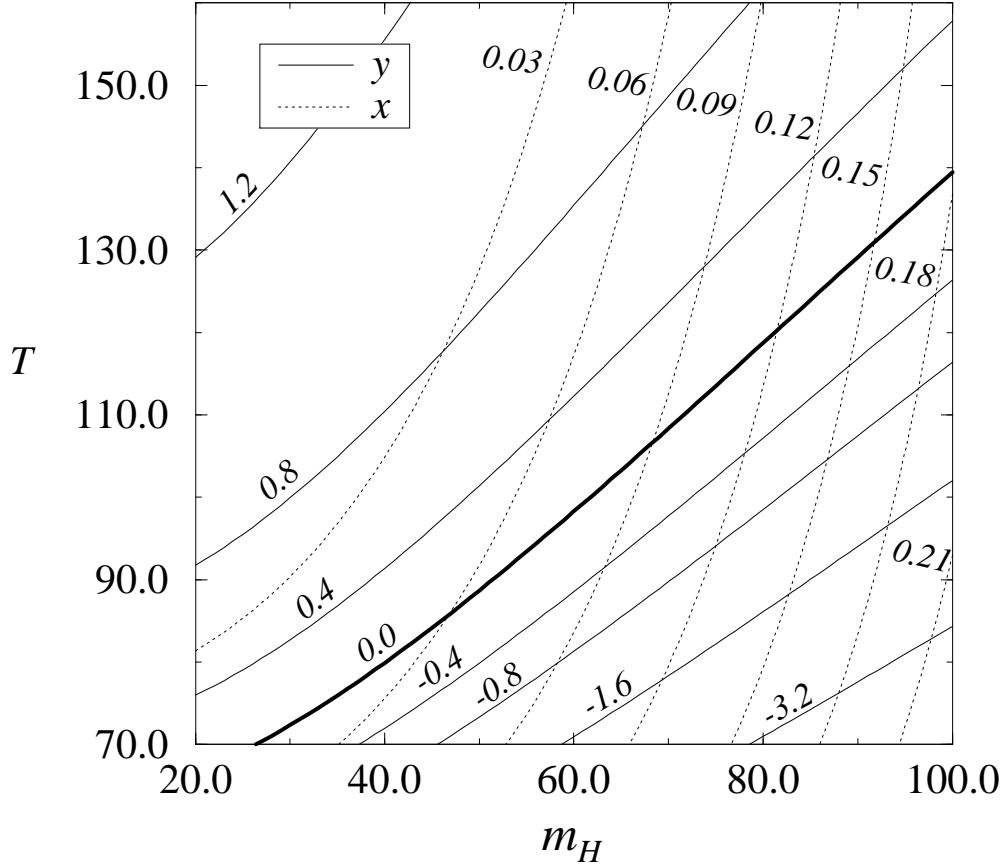}

\caption[a]{\protect
The dimensionless dynamical parameters
$x=\bar{\lambda}_3/\bar{g}_3^2$ and
$y=\bar{m}_3^2(\bar{g}_3^2)/\bar{g}_3^4$
of the 3d
SU(2)+Higgs theory as a function of $m_H$ and $T$.
The parameter $x$ depends on temperature only through
logarithmic 1-loop corrections.
The tree-level value for
the critical temperature~$T_c$ is
given by the line $y=0$,
and the true $T_c$
is rather close to this line.}
\label{xymHT}
\end{figure}

Numerically,
the properties of the 3d SU(2)$\times$U(1)+Higgs theory relevant
for the EW phase transition can be presented
as a function of the physical parameters through a few figures.
First, put $m_Z\to m_W$ so that $g'=0$. Then the final 3d
theory has three parameters: the scale is given
by $\bar{g}_3^2$, and the dynamics is given by the two dimensionless
ratios $x=\bar{\lambda}_3/\bar{g}_3^2$,
$y=\bar{m}_3^2(\bar{g}_3^2)/\bar{g}_3^4$.
The scale $\bar{g}_3^2$ is given as a function
of $m_H$ and $T$ in Fig.~\ref{g3mHT}, and the
parameters $x$ and $y$ are given in Fig.~\ref{xymHT}.
The phase diagram of the theory with the
parameters $x$, $y$,
together with the values of the latent heat,
surface tension and correlation lengths in units of $\bar{g}_3^2$,
have been studied with lattice MC simulations
in~\cite{FKLRS}.

\begin{figure}[tbh]
\vspace*{-1cm}

\hspace*{-1.5cm}
\epsfysize=14cm
\epsffile{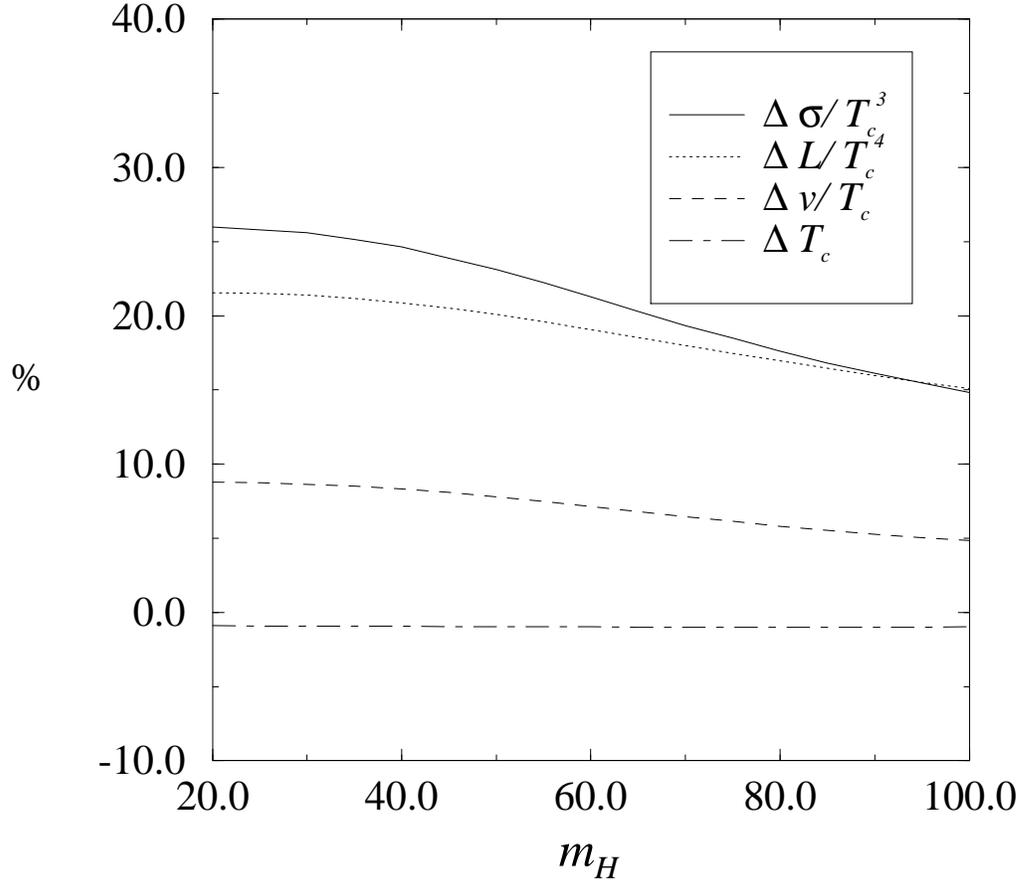}

\caption[a]{\protect
The percentual perturbative effect of the U(1)-subgroup
on the critical temperature $T_c$,
vacuum expectation value of the Higgs field $v(T_c)/T_c$,
latent heat $L/T_c^4$, and surface tension $\sigma/T_c^3$
as a function of the Higgs mass. For $v/T$, we have used the
gauge-independent definition
$v^2/T^2\equiv 2 \langle\Phi_3^\dagger\Phi_3\rangle/T$.
The surface tension is defined with the tree-level formula
$\sigma=\int d\varphi\sqrt{2 V(\varphi)}$.
The effective potential $V(\varphi)$ here is the RG-improved
2-loop effective potential for the Higgs field~\cite{FKRS1}
in the 3d SU(2)$\times$U(1)+Higgs
theory (with the convention $g'^2\sim g^3$).}
\label{U1}
\end{figure}

Finally, one has to account for
the effect of the U(1)-subgroup
on the EW phase transition. Since
there are no lattice simulations available for the 3d
SU(2)$\times$U(1)+Higgs model, the best one can do is
to estimate the effect of the U(1)-subgroup
perturbatively. In Fig.~\ref{U1}, we display the
percentual perturbative effect of the U(1)-subgroup on the
critical temperature $T_c$, the vacuum
expectation value of the Higgs field~$v$, the latent heat $L$,
and the surface tension $\sigma$ as a function of the physical Higgs mass.
Using the non-perturbative
values for the case $g'=0$ from~\cite{FKLRS}, one can then
derive results for the full Standard Model.

\subsection{The effect of higher-order operators}
\label{corrections}

In Landau gauge in the Standard Model,
the dominant 6-dimensional $\phi^6$-operators related to
the integration over the superheavy scale are
\ba
O^{(6)}_{g^2} & = & \frac{3\zeta(3)}{16\,384\pi^4}\frac{g^6\phi^6}{T^2},
\label{o6g2} \\
O^{(6)}_{g_Y^2} & = & -\frac{7\zeta(3)}{512\pi^4}\frac{g_Y^6\phi^6}{T^2}.
\label{o6gY}
\ea
These follow from eq.~\nr{hoJ} in Sec.~\ref{blocks}.
A complete list of the dominant fermionic
contributions to the other 6-dimensional
operators has been worked out in~\cite{M}.
The dominant $\phi^6$-operator related to the integration
over the heavy scale is
\be
O^{(6)}_{\rm heavy}=\frac{3\sqrt{6/5}}{10\,240\pi}\frac{g^3\phi^6}{T^2}.
\label{o6h}
\ee
The 6-dimensional operators are neglected in the effective theories
discussed in this paper, and their
importance has to be estimated.

It is rather difficult to estimate
the effect of the 6-dimensional operators
comprehensively, apart from the powercounting estimate
in Sec.~\ref{DR}. What can be done easily, though,
is an evaluation of the shift caused by the $\phi^6$-operators
in the vacuum expectation value of the Higgs field.
A generic 6-dimensional operator $O^{(6)}=c\phi^6/T^2$
produces the term $\delta V=c\varphi^6/T^2$
to the effective potential $V(\varphi)$.
Through
\be
V'(\varphi+\delta\varphi)+\delta V'(\varphi)=0,
\ee
the relative shift induced is
\be
\frac{\delta\varphi}{\varphi}=-\frac{\delta V'}{\varphi V''}
\approx  -\frac{3c}{\lambda}\frac{\varphi^2}{T^2},
\ee
where it was assumed that $V''(\varphi)\sim 2\lambda \varphi^2$.
For $m_H\sim m_W$ so that $\lambda\sim g^2/8$, the coefficients
$3c/\lambda$ in the Standard Model
following from eqs.~\nr{o6g2}, \nr{o6gY},
\nr{o6h} are
\be
\biggl(\frac{3c}{\lambda}\biggr)_{g^2}\sim 10^{-5},\quad
\biggl(\frac{3c}{\lambda}\biggr)_{g_Y^2}\sim -10^{-2},\quad
\biggl(\frac{3c}{\lambda}\biggr)_{\rm heavy}\sim 10^{-3}.
\ee
The contributions from the top quark are seen to be
dominant, and in the region where $\varphi/T\sim 1$, the effect
of the corresponding 6-dimensional operator is of the order of one
percent. Note that from the point of view of the 6-dimensional operators,
the integration over the heavy scale is relatively {\em better} than
the integration over the superheavy scale, although in terms
of powers of coupling constants, the accuracy is worse.
We conclude that the final 3d effective theory for the light
fields should give results accurate within a few percent for all
thermodynamic properties of the phase transition, like the
latent heat, the surface tension, and the correlation lengths.
For the critical temperature, the accuracy should be
an order of magnitude better. In the pure SU(2)+Higgs theory
without fermions, the accuracy of the theory with light and heavy
fields should be better than 1\% for all thermodynamic properties.

\section{Discussion}
\label{conclusions}
The set of diagrams described and computed in Section 3 is sufficient
to make a dimensional reduction of a large class of theories.
In particular, it can be used for a construction
of an effective 3d theory for different extensions of the Standard Model.
Below we will argue that in many cases the effective theory appears to
be just the SU(2)$\times$U(1)+Higgs model. We do not attempt to carry out
the necessary computations here and discuss the general strategy only.

Let us take as an example the two Higgs doublet model. The integration
over the superheavy modes gives a 3d SU(2)$\times$U(1)
theory with two Higgs doublets,
one Higgs triplet and one singlet (the last two are
the zero components of the gauge fields).
Construct now the 1-loop scalar mass matrix for the doublets and find
the temperatures
at which its eigenvalues are zero. Take the higher temperature; this is
the temperature near which the phase transition
takes place. Determine the mass of
the other scalar at this temperature. Generally, it is of the order of
$g T$, and therefore, is heavy. Integrate it out together with the
heavy triplet and singlet -- the result is the
simple SU(2)$\times$U(1) model. In the
case when both scalars are light near the critical temperature a
more complicated model, containing two scalar doublets, should be studied.
It is clear, however, that this case requires some fine tuning.
The consideration of the phase transitions in
the two Higgs doublet model on 1-loop
level can be found in~\cite{BKS,TZ}.

The same strategy is applicable to the Minimal Supersymmetric Standard Model.
If there is no breaking of colour and charge at high temperature (breaking
is possible, in principle,
since the theory contains squarks),
then all degrees of freedom, excluding those belonging to the two Higgs
doublet model, can be integrated out. We then return back to the case
considered previously. The conclusion in this case is similar to the
previous one, namely that the phase transition in MSSM can be described
by a 3d SU(2)$\times$U(1) gauge-Higgs model, at least in a part of the
parameter space. A 1-loop analysis of this theory was carried
out in~\cite{Gi,BEQ,EQZ}.

The procedure of dimensional reduction will give an infrared safe connection
between the parameters of the underlying 4d theory and those
of the 3d theory. The latter can then be studied by non-perturbative means,
such as lattice Monte Carlo simulations~\cite{KRS,FKRS,knp,IKP,FKLRS}.

\end{document}